\documentclass{aa}  

\usepackage{graphicx}
\usepackage{amsmath}
\usepackage{mathabx}
\usepackage{txfonts}
\usepackage{xcolor}
\usepackage[graphicx]{realboxes}
\usepackage{enumerate}
\usepackage[toc,page]{appendix}

\begin{document} 

\title{Tidal and general relativistic effects in rocky planet formation at the substellar mass limit using N-body simulations}
   \subtitle{}

   \author{Mariana B. S\'anchez
          \inst{1,2}\thanks{e-mail:msanchez@fcaglp.unlp.edu.ar}
          \and
          Gonzalo C. de El\'ia\inst{1,2}
          \and
          Juan Jos\'e Downes \inst{3}
          }
   \institute{Facultad de Ciencias Astron\'omicas y Geof\'isicas, Universidad Nacional de La Plata, Paseo del Bosque S/N (1900), La Plata, Argentina.\\
            \and
             Instituto de Astrof\'isica de La Plata, CCT La Plata-CONICET-UNLP, Paseo del Bosque S/N (1900), La Plata, Argentina.\\
             \and
             Centro Universitario Regional del Este, Universidad de la Rep\'ublica, AP 264, Rocha 27000, Uruguay.\\
             }
  \date{}

 
\abstract
{Recent observational results show that very low mass
stars and brown dwarfs are able to host close-in rocky
planets. Low-mass stars are the most abundant stars in
the Galaxy, and the formation efficiency of their planetary
systems is relevant in the computation of a global
probability of finding Earth-like planets inside
habitable zones. Tidal forces and relativistic effects
are relevant in the latest dynamical evolution of planets around low-mass
stars, and their effect on the planetary formation
efficiency still needs to be addressed.}
{Our goal is to evaluate the impact of tidal forces and
relativistic effects on the formation of rocky planets
around a star close to the substellar mass limit in
terms of the resulting planetary architectures and its
distribution according to the corresponding evolving
habitable zone.}
{We performed a set of $N$-body simulations spanning the first 
100~Myr of the evolution of two systems composed of 
224 embryos with a total mass 0.25M$_\oplus$ and 74 embryos with
a total mass 3 M$_\oplus$ around a central object of 0.08~M$_\odot$. 
For these two scenarios we compared
the planetary architectures that result from simulations
that are purely gravitational with those from simulations
that include the early contraction and spin-up of the
central object, the distortions and dissipation tidal
terms, and general relativistic effects.}
{We found that including these effects allows the formation and survival of a close-in ($r<0.07$~au) population of rocky planets 
with masses in the range $0.001<m/\mathrm{M_\oplus}<0.02$ in all the simulations of the less massive 
scenario, and a close-in population with masses $m \sim 0.35$~M$_\oplus$ in just a few of the simulations of the more massive scenario. The surviving close-in bodies suffered more collisions 
during the integration time of the simulations. These collisions play an important role in their 
final masses. However, all of these bodies conserved their initial amount of water in mass throughout the integration time.}
{The incorporation of tidal and general relativistic effects allows the formation of an in situ close-in population located in the habitable zone of the system. This means that both effects are relevant during the 
formation of rocky planets and their early evolution around stars close to the substellar mass limit, in particular when low-mass planetary embryos are involved.}


\titlerunning{Rocky planet formation at the substellar mass limit}
   \maketitle
%

\section{Introduction}

During the past decades several observational and theoretical results have suggested that the 
formation of rocky planets is a common process around stars of different 
masses \cite[e.g.,][]{Cumming2008,Mordasini2009,Howard2013,Ronco2017}. 
In particular, observations and modeling have proven the existence and formation 
of rocky planets around very low mass stars (VLMS) and brown dwarfs (BDs)
\citep[e.g.,][]{Payne2007,Raymond2007,Gillon2017}. These achievements are relevant 
because VLMS are the most abundant stars in the Galaxy and together with BDs are 
within the closest solar neighbors \citep[e.g.,][]{Padoan2004,Henry2004,Bastian2010}. 
This allows for surveys of rocky planets even in habitable zones, around numerous 
stellar samples, and through different observational techniques. This could be 
a crucial observational test of the processes driving the planet formation
as suggested by theoretical modeling.

Although the detection of rocky planets around BDs is still challenging,
some systems have been discovered \citep[e.g.,][]{Kubas2012,Gillon2017,Grimm2018,Z2019}.
Using photometry from \textit{Spitzer}, \citet{He2017}
reported an occurrence rate of $\sim87\%$ of planets with radius  $0.75<R/\mathrm{R_\oplus}<1.25$ 
and orbital periods  $1.7<P/\mathrm{days}<1000$ around a sample of 44 BDs.
From the M-type low-mass stars monitored by the \textit{Kepler} mission, \citet{Mulders2015b}
found that planets around VLMS are located close to their host stars, having 
an occurrence rate of small planets $(1<R/\mathrm{R_\oplus}<3),$ which is $\text{three}$ to $\text{four}$ 
times higher than for Sun-like stars, while \citet{Hardegree2019} estimated  
a mean number of $1.19$ planets per mid-type M dwarf
with radius $0.5<R/\mathrm{R_\oplus}<2.5$ and orbital period $0.5<P/\mathrm{days}<10$. 
The current \textit{SPHERE} (Spectro-Polarimetric High-contrast Exoplanet REsearch) together with \textit{ESPRESSO} (Echelle SPectrograph for Rocky Exoplanet and Stable Spectroscopic Observations) are already detecting Earth-sized planets around G, 
K, and M dwarfs \citep{Lovis2017,Hojj2019}, and \textit{CARMENES} (Calar Alto High-Resolution Search for M Dwarfs with Exo-earths with a Near-infrared Echelle Spectrograph) is searching for exoplanets around M dwarfs and 
already found two Earth-mass planets around an M7 BD \citep{Z2019}. 
The ongoing and upcoming transit searches such
as \textit{TESS} (Transiting Exoplanet Survey Satellite) and \textit{PLATO} (PLAnetary Transits and Oscillations of stars)
are expected to find most of the nearest transiting systems in the next years \citep{Barclay2018,Ragazzoni2016},
opening the new era of atmospheric characterization of terrestrial-sized planets.

Current observations suggest that there does not seem to be a discontinuity in 
the general properties of the circumstellar disks around VLMS and BDs
\citep[e.g.,][]{luhman2012}. In particular, the dust growth to millimeter 
and centimeters sizes on the disk mid-plane of BDs is similar to the growth in 
VLMS disks, as has been inferred for a few BDs that have been observed with ALMA (Atacama Large Milimeter Array) and 
CARMA (Combined Array for Research in Milimeter-wave Astronomy) \citep{Ricci2012,Ricci2013,Ricci2014}. This suggests that similar processes in the evolution of the disk might also take place 
at either side of the substellar mass limit.

\citet{Payne2007} investigated planet formation around low-mass objects 
using a standard core accretion model. They found that the formation 
of Earth-like planets is possible even around BDs, and that 
planets with masses up to 5~M$_\oplus$ can be formed. They 
reported that the mass-distribution of the resulting planets is strongly 
correlated with the disk masses. In particular, if the BD has a disk 
mass of about a few Jupiter masses, then only $10\%$ of the BDs 
might host planets with masses exceeding 0.3~M$_\oplus$. 
Through dynamical simulations of terrestrial planet formation from 
planetary embryos, \citet{Raymond2007} found that the masses of planets located inside the habitable zone decrease while the mass of the host star decreases. The authors found small and dry planets 
around low-mass stars. Using $N$-body simulations with diverse water-mass 
fractions for objects beyond the snow line, 
\citet{Ciesla2015} found both dry and water-rich planets close to low-mass stars. 
By studying planet formation around different host stars, their 
simulations predict that a greater
number of compact and low-mass planets are located around
low-mass stars, while higher mass stars will be hosting fewer and
more massive planets. Recently, \citet{Coleman2019} studied 
planet formation around low-mass stars similar to Trappist-1 and considered 
different accretion scenarios. They found water-rich rocky planets with periods
$0.5<P/\mathrm{days}<1000$. Furthermore, \citet{Miguel2019} studied planet formation around BDs and low-mass stars using a population 
synthesis code and found planets with a high ice-to-rock ratio.

It is still debated how the formation and evolution of rocky planetary
systems around low-mass objects needs to be treated. Planets around 
these objects are thought to form close to them in a region where tides 
are very strong and lead to significant orbital changes
\citep{Papaloizou2010,Barnes2010,Heller2010}.
That 
BDs as well as VLMS collapse and spin up with time in their first 100 Myr allows the population of close-in 
tidally locked bodies around them to experience many different dynamical 
evolutions \citep{Bolmont2011}. \citet{Bolmont2013,Bolmont2015} 
showed the importance 
of incorporating tidal effects and general relativity as well as the effect of
rotation-induced flattering in the dynamical and tidal evolution of multi-planetary 
systems, particularly those with close-in bodies.

In this work we incorporate these effects in a set of $N$-body simulations to study 
rocky planet formation from a sample of embryos around an object with a mass of 0.08~M$_\odot$ close to the substellar mass limit
in a  period of 100~Myr in order to improve predictions 
for planetary system architectures.
We evaluate how relevant tidal and relativistic effects 
are during rocky planet formation around an object at the substellar mass limit. 
Our aim is to estimate dynamical properties of the resulting population of close-in bodies and the efficiency of forming planets that remain 
in the habitable zone of the system.

In Section \ref{sec:vlmsbdcollapserotation} we briefly describe
the early gravitational collapse and rotation rates of VLMS and BDs.
In Section \ref{sec:habzone} we model the 
habitable zone around a star close to the substellar mass limit.
In Section \ref{sec:disk} we describe the
protoplanetary disk model that is based on observations.
In Section \ref{sec:numodel} we explain the numerical method we used to 
include tidal forces and general relativity effects in the $N$-body 
simulations. In Section \ref{sec:results} we describe the 
resulting planetary systems, and in Section \ref{sec:discussions} 
we finally summarize our conclusions and future works.

\section{Collapse and rotation of young VLMS and BD}
\label{sec:vlmsbdcollapserotation}

The BDs are stellar structures that are unable to reach the necessary 
core temperatures and densities to sustain stable hydrogen 
fusion. For solar metallicity, the substellar mass limit that 
separates BDs from VLMS is 0.072~M$_\odot$ \citep{Baraffe2015}. During the first $\sim100$~Myr of their evolution, BDs and VLMS 
share several properties such as circumstellar 
disks at different evolutionary stages \citep[e.g.,][]{luhman2012},
the formation of planets around them, their gravitational collapse,
and the subsequent increase in their rotation rates with time.
These last two properties are particularly relevant in our modeling.

In the case of VLMS, the collapse stops when they reach the main 
sequence at ages of $\sim100$~Myr. Brown dwarfs continue to collapse
slowly toward to a degenerate structure that is unable to sustain stable 
hydrogen fusion \citep[e.g.,][]{Kumar1963,Hayashi1963}. 

Available observations show that the mean rotational periods of BD are shorter 
than those of VLMS. This has been interpreted as indicating that the magnetic braking on the early spin-up in the 
substellar mass domain is inefficient \citep{Mohanty2002,Scholz2015}.
The spin rate of the main object sets the position of 
the corotation radius 
$r_\textrm{corot}$ , which is the mid-plane orbital distance at which the mean 
orbital velocity $n$ of a planet is equal to the rotational velocity
$\Omega_{\star}$ of the central object. In the case of a VLMS and BD, 
$r_\textrm{corot}$ shrinks while the object spins up. This behavior is essential for the treatment of close-in bodies because their orbital
evolution depends on the initial eccentricity $e$ and semimajor axis $a$ with respect to the location of $r_{\textrm{corot}}$. 

\section{Modeling the habitable zone}
\label{sec:habzone}

The classical habitable zone (CHZ) is the circular region around a 
single star or a multiple star system in which a rocky planet can 
retain liquid water on its surface \citep{Kasting1993}. The CHZ definition assumes that the most important greenhouse gases for 
habitable planets orbiting main-sequence stars are CO$_{2}$ and 
H$_{2}$O. This assumption extends the idea that the long-term 
($\sim$1~Myr) carbonate-silicate cycle on Earth acts as a planetary
thermostat that regulates the surface temperature 
\citep{Watson1981,Walker1981,Kasting1988} toward potentially habitable
exoplanets.

\begin{figure}
    \centering
    \includegraphics[width=9cm]{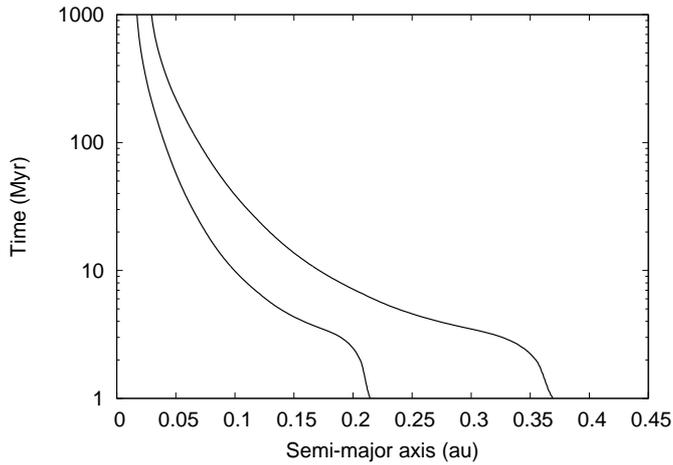}
    \caption{Evolution of the habitable zone around a star of 
    0.08~M$_\odot$ in a period of 1~Gyr.}
    \label{fig:HZ_models}
\end{figure}
 

To study potentially habitable exoplanets around a nonsolar-type star, it is necessary to take the relation between the albedo and the effective temperature of the central star into account. By extrapolating the cases studied by \citet{Kasting1993}, the inner and outer limits of the isolated 
habitable zone (IHZ) for stars with effective temperatures in between $3700 < T_{\textrm{eff}}/\mathrm{K} < 7200$ can be calculated as in \citet{Selsis2007} by 

\begin{equation}
     l_{\mathrm{int}} = \left(l_{\mathrm{in},\sun}-a_{\mathrm{in}}T_{\star}-b_{\mathrm{in}}T_{\star}^{2}\right)\left(\frac{L_{\star}}{\mathrm{L_{\sun}}}\right)^{\frac{1}{2}},
\end{equation}

\begin{equation}
     l_{\mathrm{out}} = \left(l_{\mathrm{out},\sun}-a_{\mathrm{out}}T_{\star}-b_{\mathrm{out}}T_{\star}^{2}\right)\left(\frac{L_{\star}}{\mathrm{L_{\sun}}}\right)^{\frac{1}{2}},
\end{equation}

where $l_{\textrm{in},\sun}=0.97$~au and
$l_{\textrm{out},\sun}=1.67$~au are the inner and outer limits of a system with a 
Sun-like star as central object, considering runaway greenhouse and maximum
greenhouse values, respectively \citep{Kopparapu2013,Kopparapu2013erratum}, and
$a_{\textrm{in}} = 2.7619 \times 10^{-5}\ \ \mathrm{au K^{-1}}$,
$b_{\mathrm{in}}= 3.8095 \times 10^{-9}\ \ \mathrm{auK^{-2}}$,
$a_{\mathrm{out}}=1.3786 \times 10^{-4}\ \ \mathrm{auK^{-1}}$ , and
$b_{\mathrm{out}}=1.4286 \times 10^{-9}\ \ \mathrm{auK^{-2}}$ are 
empirically determined constants;  L$_\odot$ and $L_\star$ are the luminosity of the Sun and 
the considered star, and the temperature of the star is $T_\star=T_{\textrm{eff}}-5700K,$ where $T_{\textrm{eff}}$ 
can be expressed by

\begin{equation}
  T_{\mathrm{eff}}=\left(\frac{L_{\star}}{4\pi\sigma R_{\star}^{2}}\right)^{\frac{1}{4}},
\end{equation}  

with R$_\star$ the radius of the central object and $\sigma$  the Stefan-Boltzmann constant. 
As in \citet{Barnes2013}, who studied habitable planets around BDs, we extended the calculation of
the IHZ to the substellar mass limit. In this work $l_{\textrm{in}}$ and $l_{\textrm{out}}$
therefore represent the inner and outer limits of the IHZ around a 0.08~M$_\odot$ object. We
used $R_\star$, $L_\star$ , and $T_{\textrm{eff}}$ as a
function of time as predicted by the models of 
\citet{Baraffe2015}\footnote{http://perso.ens-lyon.fr/isabelle.baraffe/BHAC15dir/}. 
In Fig.~\ref{fig:HZ_models} we show the evolution of the IHZ of an object of 0.08~M$_\odot$. 
While the object is evolving, $L_\star$, 
$R_\star$ , and $T_{\textrm{eff}}$ 
decrease with time and the location of the IHZ becomes narrower and closer 
to the substellar object. When the central object is 1~Myr old,
$l_{\textrm{in}} = 0.214$~au and $l_{\textrm{out}} = 0.369$~au, while at 1~Gyr, 
$l_{\textrm{in}} = 0.017$~au and $l_{\textrm{out}} = 0.029$~au. It is worth noting that the IHZ 
is located more than ten times closer to the central object after this time. 

\section{Protoplanetary disk}
\label{sec:disk}

In this section we describe the protoplanetary disk model that we adopted 
for the 0.08~M$_\odot$ central object. We calculate the dust species that might survive inside
our region of study to determine the material that is avialable for the formation of larger 
structures such as protoplanetary embryos. We distinguish two scenarios, motivated 
by the diversity of disk masses and the observed distribution of the exoplanet mass around low-mass objects. Finally, we compare our model with others that have been used by different authors.


\subsection{Region of study}

Our region of study is defined between an inner radius $r_{\textrm{init}} = 0.015$~au and an outer radius of $r_{\textrm{final}} = 1$~au. The inner radius was selected by considering that tidal effects allow a planet located at this distance to survive without colliding with the central object for certain values of its eccentricity \citep{Bolmont2011}. The outer radius was defined in a way that the habitable zone and an outer region of water-rich embryos are contained in our region of study.

\begin{figure}[htbp]
\includegraphics[width=9cm]{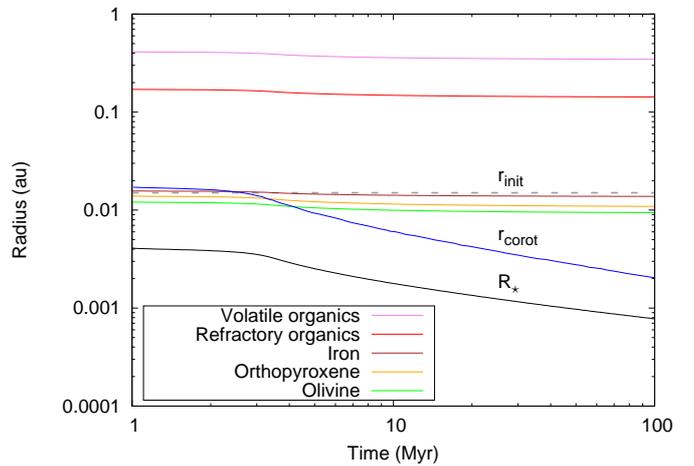}
\caption[]{Sublimation radii for different dust grain species and comparison with the corotation $r_{\textrm{corot}}$, the stellar 
radius $R_\star$ , and the inner radii of our region of study 
$r_{\textrm{init}}$.}
\label{fig:sublimation_radii}
\end{figure}

We evaluated the consistency of the selection of  $r_{\textrm{init}}$ with the sublimation radius for different dust species by analyzing those species that could survive inside our 
region of study. We computed sublimation radii for a variety of 
species using the model of \citet{Kobayashi2012}. Sublimation 
temperatures were estimated according to \citet{Pollack1994}.
In Fig.~\ref{fig:sublimation_radii} 
we show the variation of the sublimation radii with time for different 
species, compared with the corotation radii $r_{\textrm{corot}}$, the inner 
radii $r_{\textrm{init}}$ of the region, and the radius of the stellar object 
$R_\star$. Components such as iron and volatile and refractory organics could 
survive during the first million years inside our region of study.  

\subsection{Protoplanetary disk model}
\label{sec:protodisk}

The parameter that determines the distribution of the material within the disk 
is the surface density $\Sigma$. Based on physical models of viscous accretion disks \citep[see][]{Lynden-Bell1974,Hartmann1998}, we adopted the dust surface density profile 
given by

\begin{equation}
\label{eq:sigmadust}
\Sigma_{\textrm{s}}(r)=\Sigma_{0\textrm{s}}\eta_{\textrm{ice}}\left(\frac{r}{r_{\textrm{c}}}\right)^{-\gamma}\textrm{e}^{{-(r/r_{\textrm{c}})}^{2-\gamma}}.
\end{equation}
This profile is commonly used to interpret observational results 
in a wide range of stellar masses down to the substellar mass regime 
\citep[e.g.,][]{Andrews2009,Andrews2010,Guilloteau2011,Testi2016}. The value $r$
represents the radial coordinate in the mid-plane of the disk, $r_\textrm{c}$ is the
characteristic radius of the disk, $\gamma$ is the factor that determines the
density gradient, and $\eta_\textrm{ice}$ represents the increase in the amount 
of solid material due to the condensation of water beyond the 
snow line $r=r_\textrm{ice}$. For large samples of stars, \citet{Andrews2009,Andrews2010b} found that the factor $\gamma$ can take values between 1.1, and the mean value is 0.9. On the other hand, using a different technique, \citet{Isella2009} found values for $\gamma$ between -0.8 and 0.8, with a mean value of 0.1. For BD and VLMS, the lower and upper bounds for $\gamma$ are -1.4 and 1.4, respectively, and the mean value is close to 1 \citep{Testi2016}. We took $r_\mathrm{c} = 15$~au 
and $\gamma=1,$ which are consistent with the latest observations of disks 
around BDs and VLMS \citep{Ricci2012,Ricci2013,Ricci2014,Testi2016,Hendler2017}. We fixed the location 
of the snow line at $r_\textrm{ice}=0.42$~au (see Appendix \ref{sec:Apendix}). Following \citet{Lodders2003}, we propose that inside the snow line 
$\eta_\textrm{ice}=1$ and beyond the snow line $\eta_\textrm{ice}=2$. This jump of a factor 2 in the solid surface density profile is related to the water gradient distribution. Thus, 
we considered that bodies beyond $r_\textrm{ice}$ present $50\%$ of the water in mass,
while bodies inside $r_\textrm{ice}$ have just $0.01\%$ water in mass. This small percentage of water for bodies inside the snow line is given considering that the inner region was affected by water-rich embryos from beyond the snow line during the gaseous phase that is related to the evolution of the disk. The water distribution was 
assigned to each body at the beginning of our simulations. The highest initial
percentage of water in mass determines the value of the maximum percentage of water in mass that a resulting planet could have given the fact that the $N$-body code treats the collisions as perfectly inelastic ones, so that bodies conserve their mass and amount of water in mass in each collision.

By assuming an axial symmetric distribution for the solid material, we can 
express the dust mass of the disk  $M_{\textrm{dust}}$ by

\begin{equation} 
\label{eq:mdust}
M_{\textrm{dust}} = \int_{0}^{\infty} 2\pi r \Sigma_{\textrm{s}}(r) dr.
\end{equation}
Solving  Eq.~\eqref{eq:mdust} means solving two integrals 
because of the jump in the content of water in the disk given by $\eta_{\textrm{ice}}$ 
at $r_{\textrm{ice}}$.  Thus we can estimate the normalization constant for the 
solid component of the disk $\Sigma_{0s}$ for a given value of the solid 
mass in the disk. 


\subsection{Twofold parameterization of the disk density}
\label{sec:twofolddensity}

As discussed by \citet{Manara2018}, there is reliable observational evidence
that protoplanetary disks are less massive than the known exoplanet
populations. The authors suggested two mechanisms for this 
discrepancy in mass: an early formation  of  planetary  cores  at  ages 
$<0.1-1$~Myr when disks may be more massive, and replenishment of disks 
by fresh material from the environment during their lifetimes. 
In order to consider the current uncertainties 
in estimating disk masses, we made the
disk parameterization of the surface density
profile from Eq. \eqref{eq:sigmadust}
for two distinct values of $M_{\textrm{dust}}$. We
refer to these two cases as the following 
\emph{\textup{mass scenarios}}:

\begin{itemize}
    \item \textbf{\textup{The disk scenario (S1)}} is based on the 
    latest observational results on the masses of dust in protoplanetary 
    disks. We assumed $M_{\textrm{dust}} = 9 \times 10^{-6}$~M$_\odot$ ($\sim$ 3~M$_\oplus$) 
    from the average of the dust masses obtained from observations of BDs and 
    VLMS made with ALMA (see references in Section~\ref{sec:protodisk}). If we were to assume 
    a gas-to-dust ratio of 100:1, this would be equivalent to taking $M_{\textrm{disk}} =1.1\%$~M$_\star$.
    \newline
    \item \textbf{The planetary systems scenario (S2)} is based on 
    the observational results on the masses of exoplanetary systems. We assumed 
    $M_{\textrm{dust}} =  9 \times 10^{-5}$~M$_\odot$ ($\sim30$~M$_\oplus$) regarding the 
    current terrestrial exoplanet detection around BDs
    \citep[e.g.,][]{Kubas2012,Gillon2017,Grimm2018}. If we were to assume a gas-to-dust 
    radio of 100:1, this would be equivalent to taking $M_{\textrm{disk}} =11\%$~M$_{\star}$. In 
    this case, we increased the percentage of the mass in order to extend the
    solid material in the disk that is available in our region of study to form rocky planets.
\end{itemize}

\subsection{Contrasting $\Sigma$ parameterization}

\begin{figure}[htbp]
\includegraphics[width=9cm]{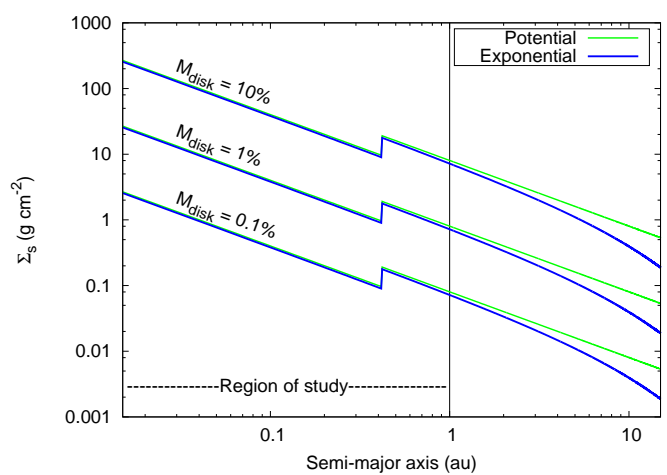}
\caption[]{Power law (green line) and exponential tapered (blue line) 
surface density profiles in the protoplanetary-planetary disk for a total disk mass $M_{\textrm{disk}}$ of $0.1\%$, $1\%,$ and $10\%$ of the mass of 
the central object of 0.08~M$_\odot$. The region of study ($0.015<r/\mathrm{au}<1$) 
is indicated.}
\label{fig:densprofiles_varios}
\end{figure}

 Many authors have also proposed a power-law surface density 
 profile to model protoplanetary disks \citep[e.g.,][]{Ciesla2015,Testi2016}.
 We therefore compared the model proposed in this work (Eq. \ref{eq:sigmadust}) 
 with a power-law density profile, 
 
\begin{equation} 
\label{eq:powlawdensprof}
\Sigma_{\textrm{sp}}(r) = \Sigma_{\textrm{sp}0}\eta_{\textrm{ice}}\left(\frac{r}{r_0}\right)^{-p}
,\end{equation}
 where $r_0$ and $p$ are equivalent to $r_c$ and $\gamma$ in the
 exponentially tapered density profile. In Fig.~\ref{fig:densprofiles_varios} 
 we show the comparison of the two density profiles considering the same initial
 parameters as we chose to describe $\Sigma_\textrm{s}(r)$ (see Section
 \ref{sec:protodisk}). As an example, we selected three different disk masses
 $M_{\textrm{disk}}$: $0.1\%$, $1\%,$ and $10\%$ of the mass of the central 
 object, and we assumed a gas-to-dust ratio of $100:1$. The power-law profiles do not 
 show significant differences with the exponentially tapered density 
 profiles within our region of study.

\section{Numerical model}
\label{sec:numodel}

In this section we describe the treatment of planet formation around an object of 0.08~M$_\odot$ by including a protoplanetary embryo distribution 
 that interacts with the main object. We developed a set of $N$-body simulations 
with the well-known $\textsc{Mercury}$ code \citep{Chambers1999} by incorporating
tidal and general relativistic acceleration corrections as external forces. 
Thus the dynamical evolution of protoplanetary embryos was affected not only 
by gravitational interactions between them and with the star, but also by 
tidal distortions and dissipation, as well as by general relativistic effects. 
The stellar contraction and rotational period evolution was included in the code as 
well as a fixed pseudo-synchronization period for protoplanetary embryos during 
the 100~Myr integration time of our simulations.

\subsection{Tidal model}
\label{sec:tidalmodel}

We followed the equilibrium tide model \citep{Hut1981,Eggleton1998} that was rederivated by \citet{Bolmont2011}, which considers both the tide raised by the BD on the planet and by the planet on the BD in the orbital evolution of planetary systems. It also takes into account the spin-up and contraction of the BD. The authors followed the constant time-lag model and assumed constant internal dissipation for the BD and the planets involved.

Following the equilibrium tidal model, we incorporated tidal distortions and dissipation terms, considering the 
tide raised by the star on each protoplanetary embryo and by each protoplanetary 
embryo on the star and neglected the tide between embryos. Tidal interactions produce deformations on the
bodies that in a heliocentric reference frame lead to precession of the  argument of periastron $\omega$ 
and a decay in semimajor axis $a$ and eccentricity $e$ , which can be interpreted as distortions and dissipation terms, respectively.\\

The correction in the acceleration of each protoplanetary embryo 
produced by the tidal distortion term was taken from \citet{Hut1981} 
(for an explicit expression, see \citet{Beauge2012}) and is given by

\begin{equation}
  \textbf{f}_\omega = -3\frac{\mu}{r^8}\left[k_{2,\star}\left(\frac{M_\mathrm{p}}{M_\star}\right)R_\star^5 + k_{2,\mathrm{p}}\left(\frac{M_\star}{M_\mathrm{p}}\right)R_\mathrm{p}^5\right] \textbf{r},
\end{equation}  
where \textbf{r} is the position vector of the embryo with respect to the central 
object, $k_{2,\star}=0.307$ and $k_{2,\textrm{p}}=0.305$ are the potential Love numbers of 
degree 2 of the star and the embryo, respectively \citep{Bolmont2015}, 
$\mu=G(M_\star + M_\textrm{p})$, $G$ is the gravitational constant, and $M_\star$, $R_\star$, 
$M_\textrm{p}$ , and $R_\textrm{p}$ are the masses and radius of the star and the
protoplanetary embryo under the approximation that these objects can instantaneously 
adjust their equilibrium shapes to the tidal force and considering only distoritions up to the 
second-order harmonic \citep{Darwin1908}. 

The evolution of $R_\star$ was taken from the models of \citet{Baraffe2015}, and 
the value of $R_\textrm{p}$ of each protoplanetary embryo was calculated by considering each 
of them as a spherical body with a fixed volume density $\rho=5~\mathrm{gr/cm^3}$.

The timescale associated with the tidal dissipation term was calculated based on 
the work of \citet{Sterne1939} by considering the stellar and embryo 
tide, and it is given by  

\begin{equation}
  t_{\textrm{tide}} \sim  \frac{2 \pi a^{5}}{7.5nf(e)}\left(\frac{M_{\star}M_{\mathrm{p}}}{k_{2,\star}M_{\mathrm{p}}^{2}R_{\star}^{5} + k_{2,\mathrm{p}}M_{\star}^{2}R_{\mathrm{p}}^{5}}\right),\\
\label{eq:jordantimescale}
\end{equation}
with $f(e) = (1-e^2)^{-5}[1 + (3/2)e^2 + (1/8)e^4]$.\\
\\
\indent The acceleration correction of each protoplanetary embryo induced by the tidal 
dissipation term, which produces $a$ and $e$ 
decay, was obtained from \citet{Eggleton1998}. After 
some algebra, this equals the expression from \citet{Beauge2012},

\begin{align*}
\textbf{f}_{\textrm{ae}} = & -3\frac{\mu}{r^{10}} \left[
  \frac{M_{\textrm{p}}}{M_\star} k_{2,\star} \Delta \mathrm{t}_\star R_\star^{5}\left(2\textbf{r}(\textbf{r} \cdot \textbf{v}) + r^{2}(\textbf{r} \times \Omega_\star + \textbf{v})\right)\right]
\end{align*}
\begin{equation}
-3\frac{\mu}{r^{10}} \left[\frac{M_\star}{M_{\textrm{p}}}k_{2,\textrm{p}} \Delta \mathrm{t}_{\textrm{p}} R_\textrm{p}^{5} 
       \left(2\textbf{r}(\textbf{r}\cdot\textbf{v}) + r^{2}(\textbf{r} \times \Omega_\textrm{p} + \textbf{v})\right)\right],
\end{equation}
where $\textbf{v}$ is the velocity vector of the embryo, and $\Delta \mathrm{t}_\star$ 
and $\Delta \mathrm{t}_\mathrm{p}$ are the time-lag model constants 
for the star and the protoplanetary embryo, respectively. The factors 
$k_{2,\star}\Delta \mathrm{t}_\star$ , and $k_{2,\mathrm{p}}\Delta \mathrm{t}_\mathrm{p}$ 
are related to the dissipation factors by

\begin{equation}
  k_{2,\mathrm{p}} \Delta \mathrm{t} _\mathrm{p} =  \frac{3R_\mathrm{p}^5\sigma_\mathrm{p}}{2G} \\
   k_{2,\star} \Delta \mathrm{t}_{\star} = \frac{3R_{\star}^5\sigma_{\star}}{2G}
,\end{equation}  
with the dissipation factor for each protoplanetary embryo 
$\sigma_\mathrm{p}=8.577\times10^{-50}\mathrm{g^{-1} cm^{-2} s^{-1}}$, 
the same dissipation factor as estimated for the Earth \citep{Neron1997}, 
and the dissipation factor of the central object is 
$\sigma_\star=2.006\times10^{-60}\mathrm{g^{-1} cm^{-2} s^{-1}}$ 
\citep{Hansen2010}. 

In the constant time-lag model, in which the time-lag constant $\Delta \mathrm{t}_\star$ is independent of the tidal frequency, the rotation of the companions leads to pseudo-synchronization \citep{Hut1981,Eggleton1998}. In preliminary simulations, \citet{Leconte2010,Bolmont2011,Bolmont2013} verified that a planet reaches the pseudo-synchronization very quickly in its evolution. For a planet, being at pseudo-synchronization means that its rotation tends to be synchronized with the orbital angular velocity at periastron, where the tidal interactions are stronger \citep{Hut1981}. As in \citet{Bolmont2011}, we fixed each protoplanetary embryo at pseudo-synchronization \citep{Hut1981} in each time-step of our simulations as

\begin{equation}
\label{eq:pseudosincro}
\Omega_\mathrm{p} = \frac{(1 + (15/2)e^2 + (45/8)e^4 + (5/16)e^6)}{(1 + 3e^2 + (3/8)e^4)(1 - e^2)^{3/2}} n,
\end{equation}
where $\Omega_{\textrm{p}}$ is the rotational velocity of the protoplanetary embryo.

If $e=0$, then the embryo is in perfect synchronization, thus $\Omega_\textrm{p} = n$ and only 
the tide of the central object remain. When $e$ is small, the tide of the main object dominates and determines the 
evolution of the embryo: if the embryo is located beyond $r_{\textrm{corot}}$, then 
$\Omega_\textrm{p} < \Omega_\star,~\frac{da}{dt} > 0,$ so that the embryo is pushed outward, 
and if it is inside, $\Omega_\textrm{p} > \Omega_\star,~\frac{da}{dt} < 0,$ so that the embryo 
is pulled inward. On the other hand, when $e$ is high, the embryo tide will
prevail. In this case, the embryo is pulled inward. This is always true because 
for a body in pseudo-synchronization, the body tide always acts to decrease the 
orbital distance \citep{Leconte2010}.

The rotational velocity $\Omega_{\star}$ of the main object was calculated following the tidal 
model proposed by \citet{Bolmont2011}, who integrated its evolution as affected by its 
contraction and the influence of orbiting planets. They calibrated their results 
with a set of observationally determined $\Omega_{\star}$ for VLMS and BDs at 
different ages from \citet{Herbst2007}. Thus the evolution of $\Omega_\star$ 
can be expressed as

\begin{equation}
\Omega_\star(t)=\Omega_\star(t_0)\left[\frac{r_\mathrm{gyr}^2(t_0)}{r_\mathrm{gyr}^2(t)}\left(\frac{R_\star(t_0)}{R_\star(t)}\right)^{2}\times exp\left(\int_{t_0}^{t}f_{\textrm{t}} dt\right)\right]
\label{eq:omega}
\end{equation}
\citep[e.g.,][]{Bolmont2011}, where $r_\mathrm{gyr}^2$ is the square of the gyration radius, defined as $r_\mathrm{gyr}^2 = \frac{I_\star}{M_\star R_\star^{2}}$ , with $I_\star$ the moment of inertia 
of the main object \citep{Hut1981}. The function $f_{\textrm{t}}$ is given by

\begin{equation}
f_\mathrm{t}=\frac{1}{\Omega_\star}\frac{d\Omega_\star}{dt}.
\end{equation}

If we were to consider $r_\mathrm{gyr}^2$ and $R_\star$ as constant values \citep{Bolmont2011}, then

\begin{equation}
\label{eq:ftide}
f_\mathrm{t}=-\frac{\gamma_\star}{t_{\mathrm{dis},\star}}\left[No1(e) - \frac{\Omega_\star}{n}No2(e)\right]    
,\end{equation}
with $\gamma_\star = \frac{h}{I_\star\Omega_\star}$, where $h$ is the orbital angular 
momentum, $t_\mathrm{dis,\star}$ is the dissipation timescale of the central object (see below), and the functions $No1(e)$ and $No2(e)$ are dependent on the eccentricity of the planetary companion, which is given by

\begin{equation}
\begin{aligned}
No1(e)&=\frac{1+(15/2)e^{2}+(45/8)e^{4}+(5/16)e^{6}}{(1-e^{2})^{13/2}}\\   
\\
No2(e)&=\frac{1+3e^{2}+(3/8)e^{4}}{(1-e^{2})^{5}}.
\end{aligned}
\end{equation}

\begin{figure}[htbp]
\centering
 \includegraphics[width=9cm]{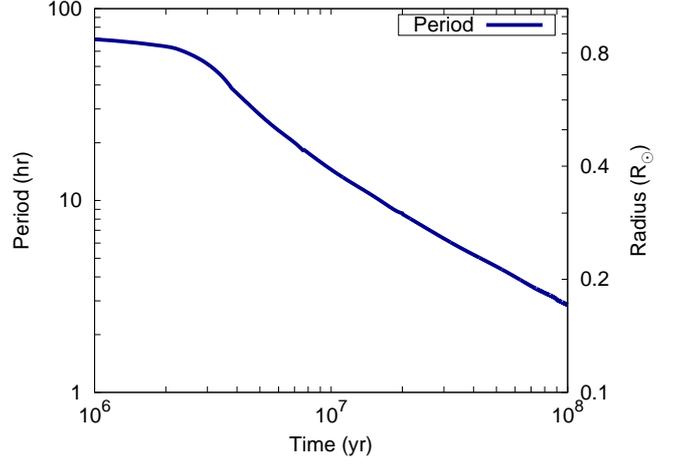}
  \caption[]{Rotation period evolution of an object of 0.08~M$_\odot$ based 
on \citet{Bolmont2011} associated with its radius contraction. The radius evolution is taken from \citet{Baraffe2015}.}
\label{fig:RandP}
\end{figure}

When only terrestrial planets are considered to orbit the host object, then $f_\textrm{t}$ 
is small and the substellar object rotation period is mainly determined by the conservation 
of angular momentum, that is, by the initial rotation period \citep{Bolmont2011}. We therefore numerically 
integrated Eq. \eqref{eq:omega} independently of the dynamics of the planetary system. We considered that the radius R$_\star$ evolves
according to the structure and atmospheric models from
\citet{Baraffe2015}, but we fixed its value for each time-step of our integration, which was small enough to be considered constant in order to simplify the integration and be able to use Eq. \eqref{eq:ftide}. We also assumed one Earth-like planet to orbit the main object with random initial values for $e$ and $a$ inside our region of study. From the different orbital elements initially given to the Earth-like planet, we verified that the evolution of $\Omega_\star$ was mainly determined by the evolution of the substellar object and was similar to the evolution achieved by \citet{Bolmont2011}. In Fig.~\ref{fig:RandP} we show the
resulting evolution of the rotational period and the 
corresponding $R_{\star}$ in a period from 1~Myr to 100~Myr.\\

The dissipation timescales for eccentric orbits are determined from the secular 
tidal evolution of $a$ and $e$ \citep[see][]{Hansen2010,Bolmont2011,Bolmont2013} 
as $t_\textrm{a}$ and $t_\textrm{e}$ , respectively, by
\begin{equation}
\begin{split}
\frac{1}{t_\textrm{a}}=\frac{1}{a} \frac{da}{dt} =  - \frac{1}{t_\textrm{dis,p}} \left[Na1(e) - \frac{\Omega_p}{n}Na2(e) \right] 
\\
- \frac{1}{t_{\mathrm{dis},\star}} \left[Na1(e) - \frac{\Omega_\star}{n}Na2(e) \right]
\end{split}
\end{equation}
\begin{equation}
\begin{split}
    \frac{1}{t_\textrm{e}}=\frac{1}{e} \frac{de}{dt}  = & - \frac{9}{2t_\textrm{dis,p}} \left[Ne1(e) - \frac{11}{18} \frac{\Omega_p}{n}Ne2(e)\right], \\
    \\
    &- \frac{9}{2t_{\mathrm{dis},\star}}\left[ Ne1(e) - \frac{11}{18} \frac{\Omega_\star}{n}Ne2(e) \right].
\end{split}    
\end{equation}
Here $t_\textrm{dis,p}$ and $t_{\textrm{dis},\star}$ are the dissipation timescales for circular orbits 
for the embryo and the main object, respectively, and $Na1$, $Na2$, $Ne1,$ and $Ne2$ 
are factors that take place in eccentric orbits and are defined by

\begin{equation}
\begin{aligned}
  t_\textrm{dis,p} &= \frac{1}{9} \frac{M_\textrm{p}}{M_\star(M_\textrm{p} + M_\star)} \frac{a^8}{R_\textrm{p}^{10}} \frac{1}{\sigma_\textrm{p}},\\
 \\
  t_{\mathrm{dis},\star} &= \frac{1}{9} \frac{M_\star}{M_\textrm{p}(M_\textrm{p} + M_\star)} \frac{a^8}{R_\star^{10}} \frac{1}{\sigma_\star},\\
  \\
    Na1(e) &= \frac{1 + (31/2)e^{2} + (255/8)e^{4} + (185/16)e^{6} + (25/64)e^{8}}{(1-e^{2})^{15/2}},\\
    \\
    Na2(e) &= \frac{1 + (15/2)e^{2} + (45/8)e^{4} + (5/16)e^{6}}{(1-e^{2})^{6}},\\
    \\
    Ne1(e) &= \frac{1 + (15/4)e^{2} + (15/8)e^{4} + (5/64)e^{6}}{(1-e^{2})^{13/2}},\\
    \\
    Ne2(e) &= \frac{1 + (3/2)e^{2} + (1/8)e^{4}}{(1-e^{2})^{5}}.
  \end{aligned}
\end{equation}

\subsection{General relativistic effect}

The important effect derived from General Relativity theory (GRT) on the dynamic of planetary systems is the precession of $\omega$ \citep{Einstein1916}. In our case, 
we considered that only the main object contributes with relevant corrections. As 
we worked in the reference frame of the star, 
the associated correction in 
the acceleration of the embryo is

\begin{equation}
  \textbf{f}_{\mathrm{GR}} = \frac{GM_\star}{r^3c^2}\left[\left(\frac{4GM_\star}{r} - \textbf{v}^2\right)\textbf{r}+4(\textbf{v}.\textbf{r})\textbf{v}\right],
  \label{eq:grav}
\end{equation}
with $c$ the speed of light. Eq. \eqref{eq:grav} was proposed by 
\citet{Anderson1975}, who worked under the parameterize post-Newtonian theories. The authors obtained a relative correction associated with two parameters $\beta$ and 
$\gamma$, which are equal to unity in the GRT case. This expression has been used in 
several works that included relativistic corrections \citep[e.g.,][]{Quinn1991,Shahid1994,Varadi2003,Benitez2008,Zanardi2018}. 
The timescale associated with the precession of the longitude of periastron is given by 

\begin{equation}
t_{\mathrm{GR}} \sim 2 \pi \frac{a^{\frac{5}{2}}c^2(1-e^2)}{3G^{\frac{3}{2}}(M_\star+M_{\textrm{p}})^{\frac{3}{2}}}.
\label{eq:GR}
\end{equation}

\subsection{Test simulations}

\begin{figure}[htbp]
\centering
 \includegraphics[width=0.45\textwidth]{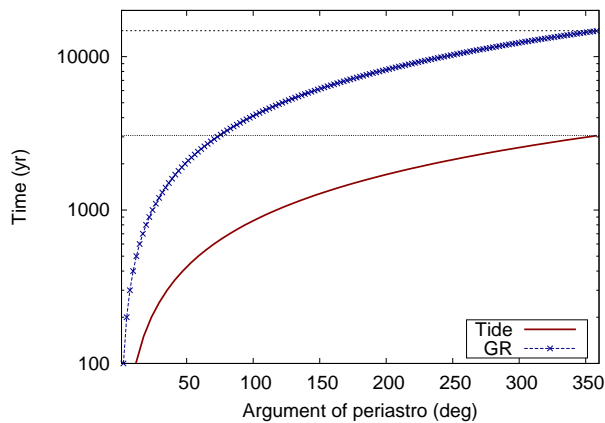} 
\caption[Apsidal precession due to tide and GR]{Apsidal precession due to 
tidal distortion (solid red line) and GR effects (dotted blue line)
in a system composed of a 1~M$_\oplus$ planet around a 0.08~M$_\odot$
BD with initial $a = 0.01$~au and $e = 0.1$. In this example, the 
argument of periastron completes an orbit in $\sim$14\,750~yr when is 
affected by GR and in $\sim$3\,060~yr when is affected by tidal distortion.}
\label{fig:perihelion}
\end{figure}

We made a set of $N$-body simulations in order to test the agreement between 
the external forces that we incorporated in the \textsc{Mercury} code and the 
timescale associated with them.
To test the precession of $\omega,$ we developed two simulations: one that included the tidal distortion term, and another that included the GR correction. In Fig.~\ref{fig:perihelion} we show the apsidal precession timescale of
a planet with 
1~M$_\oplus$ orbiting a 1~Myr substellar object of 0.08~M$_\odot$ with initial 
values $a = 0.01$~au and $e = 0.1$ for the two simulations we made. Our results show that the apsidal precession of $360^\circ$ 
is completed in 14\,750~yr and 3\,060~yr, respectively, which 
agrees with the time predicted by the timescales associated 
with each correction term in Eqs. \eqref{eq:GR} and
\eqref{eq:jordantimescale}. These timescale values depend on 
the physical parameters of the protoplanetary  embryos and the substellar object, as well as on 
the initial orbital elements. For instance, if the protoplanets are smaller 
than Earth in mass and radius, then the relativistic effect is 
more relevant than the tidal distortion regarding the precession of $\omega$.

To test the analytic expressions for the tidal dissipation with the 
timescales of $e$ and $a$ decay, we developed a $N$-body simulation 
that includes the dissipation term. Our aim was 
to compare our results with those obtained by \citet{Bolmont2011}, who
used the analytic tidal model. In their work, they represent the evolution 
of $a$, $e,$ and the rotation period of a planet of 1~M$_\oplus$ evolving around
a BD of 0.04~M$_\odot$. We 
chose as initial values $a=0.017242$~au and $e=0.744$. 
The semimajor axis we selected represents  $a_{\textrm{switch}}$, that 
is, the one that determines the 
behavior of the protoplanet and separates inward migration and crash from 
inward migration but survival of the protoplanet or outward migration. 

\begin{figure}[htbp]
\centering
  \includegraphics[width=0.48\textwidth]{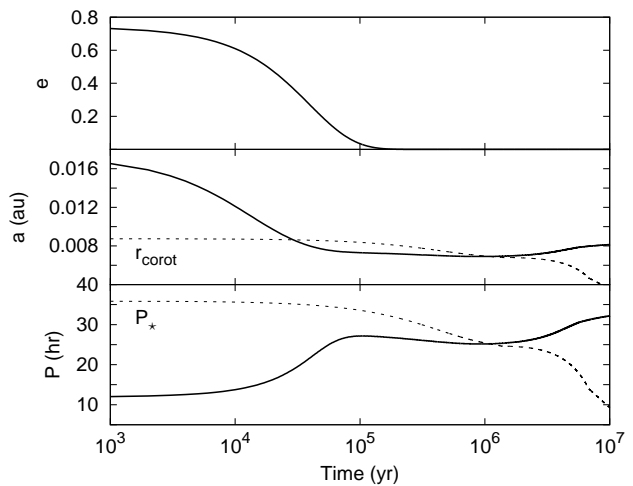}
\caption[Evolution of 1~M$_\oplus$ around a 0.04~M$_\odot$ BD]{Evolution 
of $e$, $a,$ and the pseudo-synchronization period of a
1~M$_\oplus$ around a 0.04~M$_\odot$ BD. Solid lines indicates the 
results considering the initial orbital elements from \citet{Bolmont2011}. 
The dashed lines indicate the evolution of $r_{\textrm{corot}}$ (middle panel) 
and the rotational period of the BD, $P_\star$ (bottom panel).}
\label{fig:Bolmont_example}
\end{figure}

In Fig.~\ref{fig:Bolmont_example} we show the evolution of $a$, $e,$ and 
the pseudo-synchronization period $P_\textrm{p}$ of a 1~M$_\oplus$ planet
around an evolving 0.04~M$_\odot$ BD. 
In the middle panel, we also show the $r_{\textrm{corot}}$ evolution, while in 
the bottom panel we include the rotation period of the BD, $P_{\star}$
. First, as $P_\star > P_\textrm{p}$ , the planet moves inward 
of the central object. Under this condition, even tough $a = r_\textrm{corot}$, the orbit is not circular and the planet continues to move inward. When $P_\star = P_\textrm{p}$, the orbit is circular and then $\Omega_\textrm{p} = n,$ which means that this time when 
$a = r_{\textrm{corot}}$ , the planet starts to move outward because $P_\star < P_\textrm{p}$.    

\subsection{$N$-body simulations}

We performed 20 $N$-body simulations using the modified version of the \textsc{Mercury} code. We developed 10 simulations for scenario \emph{S1} and 10 simulations for scenario
\emph{S2,} as explained in Section \ref{sec:twofolddensity}.
We also ran 10 simulations for each scenario described above using 
the original version of the \textsc{Mercury} code, without external 
forces, in order to evaluate the relevance of tidal and general relativistic 
effects.

\subsubsection{Protoplanetary embryo distributions}

For using $\textsc{Mercury}$ code, it is necessary to give physical and orbital
parameters for the protoplanetary embryos. We modeled the initial mass distributions 
of embryos as a function of the radial distance from the central object, which are 
initial conditions for our numerical simulations. The initial spatial distribution 
of protoplanetary embryos was computed following Eq.~\eqref{eq:sigmadust}, considering a distance range
$0.015<r/\textrm{au}<1$ and defining 1~Myr as the initial time. We considered that at 
this age, the gas has already been dissipated from the disk. 

Even though we are aware of the existence of a
number of BDs that still accrete gas from
their disks up to $\sim 10$~Myr
\citep[references, e.g., in][]{Pascucci2009,Downes2015}, 
incorporating the gas component is beyond 
the scope of this work, which
reproduces the BDs that are not observed to show
gas signatures at $\sim 1$~Myr, however. We calculated the mass of each protoplanetary embryo $M_\textrm{p}$ 
considering that at the initial time, they are at the end of the 
oligarchic growth stage, having accreted all the planetesimals 
in their feeding zones \citep{Kokubo2000} by

\begin{equation}
M_\textrm{p}=2\pi r\Delta r_\textrm{H}\Sigma_s(r),
\label{eq:masaemb}
\end{equation}
where $\Delta r_\textrm{H}$ is the orbital separation between two 
consecutive embryos in terms of their mutual Hill radii
$r_\textrm{H}$, with $\Delta$ an arbitrary integer number, given by

\begin{equation}
r_\textrm{H}=r\left(\frac{2M_\textrm{p}}{3 M_\star}\right)^{\frac{1}{3}}.
\label{eq:radiohill}
\end{equation}

By replacing Eqs. \eqref{eq:sigmadust} and \eqref{eq:radiohill} in 
Eq. \eqref{eq:masaemb}, we obtain an expression for the mass of each 
protoplanetary embryo as a function of the radial distance in 
the disk mid-plane $r$, which is given by

\begin{equation}
M_{\mathrm{p}}=\left(2 \pi r^{2} \Delta \Sigma_{0\textrm{s}} \eta_{\textrm{ice}} \left(\frac{2}{3 M_{\star}}\right)^{\frac{1}{3}}\left(\frac{r}{r_{\textrm{c}}}\right)^{-\gamma}\textrm{e}^{{-\left(\frac{r}{r_{\textrm{c}}}\right)}^{2-\gamma}}\right)^{\frac{3}{2}}.
\label{eq:massemb}
\end{equation}

We located our first embryo at the inner radii of our region $r_1 = 0.015$~au. 
Then we calculated its mass using Eq. \eqref{eq:massemb}. For the remaining embryos,
we propose a separation of 10 r$_{\textrm{H}}$  by fixing $\Delta=10$ \citep{Kokubo1998}.

Thus we calculated the initial positions $r_{i+1}$ 
and masses $M_{\textrm{p},i+1}$ for the embryos by 

\begin{equation}
  r_{i+1} = r_i + \Delta r_\textrm{i} \left(\frac{2M_i}{3 M_\star}\right)^{\frac{1}{3}},
  \label{eq:distancias}
\end{equation}

\begin{equation}
  M_{\textrm{p},i+1}= \left(A\left(\frac{2}{3 M_\star}\right)^{\frac{1}{3}}\left(\frac{r_{i+1}}{r_c}\right)^{-\gamma}\textrm{e}^{{-\left(\frac{r_{i+1}}{r_c}\right)}^{2-\gamma}}\right)^{\frac{3}{2}},
  \label{eq:masas} 
\end{equation}
for $i = 1,~2,~\text{etc.}$ with $A=2 \pi r_{i+1}^2\Delta\Sigma_{0\textrm{s}} \eta_\textrm{ice}$. 

\begin{figure}[htbp]
\centering
\includegraphics[width=8cm]{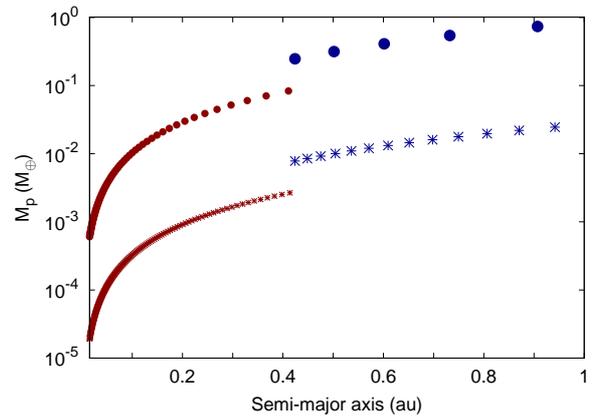}
\caption[Initial embryo distribution]{Initial embryo distributions of masses 
as a function of their initial location given by their semimajor axis for S1 (circles) and S2 (stars). 
Blue represents the water-rich population ($50\%$ water in mass), and red represents the bodies with the lowest amount of water in mass ($0.01\%$).}
\label{fig:Distribution_embryos}
\end{figure}

Using Eqs. \eqref{eq:distancias} and \eqref{eq:masas}, we derived the 
initial distributions of masses of the protoplanetary embryos as a function of 
the radial distance, which represents the semimajor axis, from the central object 
for scenarios S1 and S2. In Fig.~\ref{fig:Distribution_embryos} we illustrate 
the two distributions. S1  has a distribution of $224$ embryos with a total mass
$M_{\textrm{pT}}\sim0.25$~M$_\oplus$ located in the region of study. $\text{Two hundred and ten}$ of them are distributed in the inner region up to the snow line,
with a total mass $\sim0.06$~M$_\oplus$, while the remaining 14 embryos are distributed 
beyond the snow line and have a total mass $\sim0.19$~M$_\oplus$. 
 S2 has a distribution of $74$ embryos that are located in the region of study 
with a total mass  $M_{\textrm{pT}}\sim3$~M$_\oplus$. $\text{Sixty-nine}$ of them are distributed in the inner region 
up to the snow line, with a total mass $\sim0.72$~M$_\oplus$, 
while the remaining $5\text{}$ embryos have a total mass of $\sim2.25$~M$_\oplus$ and are 
placed beyond the snow line up to 1~au.

We considered lower values than $0.02$ for the initial eccentricities and 
$0.5^\circ$ for the initial inclinations. The  orbital elements, argument of periastron $\omega$, longitude of the ascending node $\Omega,$ and mean anomaly 
$M$ were determined randomly at the beginning of the simulations. They were between $0^\circ$ and $360^\circ$ from a uniform distribution for each protoplanetary embryo.

\subsubsection{$N$-body code: characterization}
\label{sec:proto.planet.desc}

To develop our simulations, we chose the hybrid integrator, which uses 
a second-order symplectic algorithm to treat interactions between objects with 
separations greater than $3$ Hill radii, and we selected the Bulirsch-St\"oer method to 
resolve closer encounters. 
The collisions were treated as perfectly inelastic, conserving the 
mass and the corresponding water content of protoplanetary embryos. We considered 
that a body is ejected from the system when it reaches a distance $a>100$~au.

We adopted a time step of 0.08 days, which corresponds to $1/30$~th
of the orbital period of the innermost body in the simulations. 
In order to avoid any numerical error for small-perihelion orbits, we 
assumed $R_\star=0.004$~au, which corresponds to the maximum value of the radius of the central object.

All simulations were integrated over 100~Myr, which is a standard 
time for studying the dynamical evolution of planetary systems.
Because of the stochastic nature of the 
accretion process between the protoplanets and eventually with the main object, we remark that it is necessary to carry out a set of $N$-body simulations. 
In this case, we performed ten simulations for each scenario, which required a mean CPU time of six months on 3.6~GHz processors.

\section{Results}
\label{sec:results}

In this section we present the resulting planetary systems of the simulations in scenarios S1 and S2. We compare the resulting planets from 
simulations that included tidal and GR effects with those 
from simulations
that neglected these effects to test their relevance in the formation of rocky planets. In particular, we focus our analysis on the 
population that is located close to the central object.

\subsection{Planetary architectures}
\label{sec:planet_arch}

\begin{figure*}
\centering
    \includegraphics[width=0.4\textwidth]{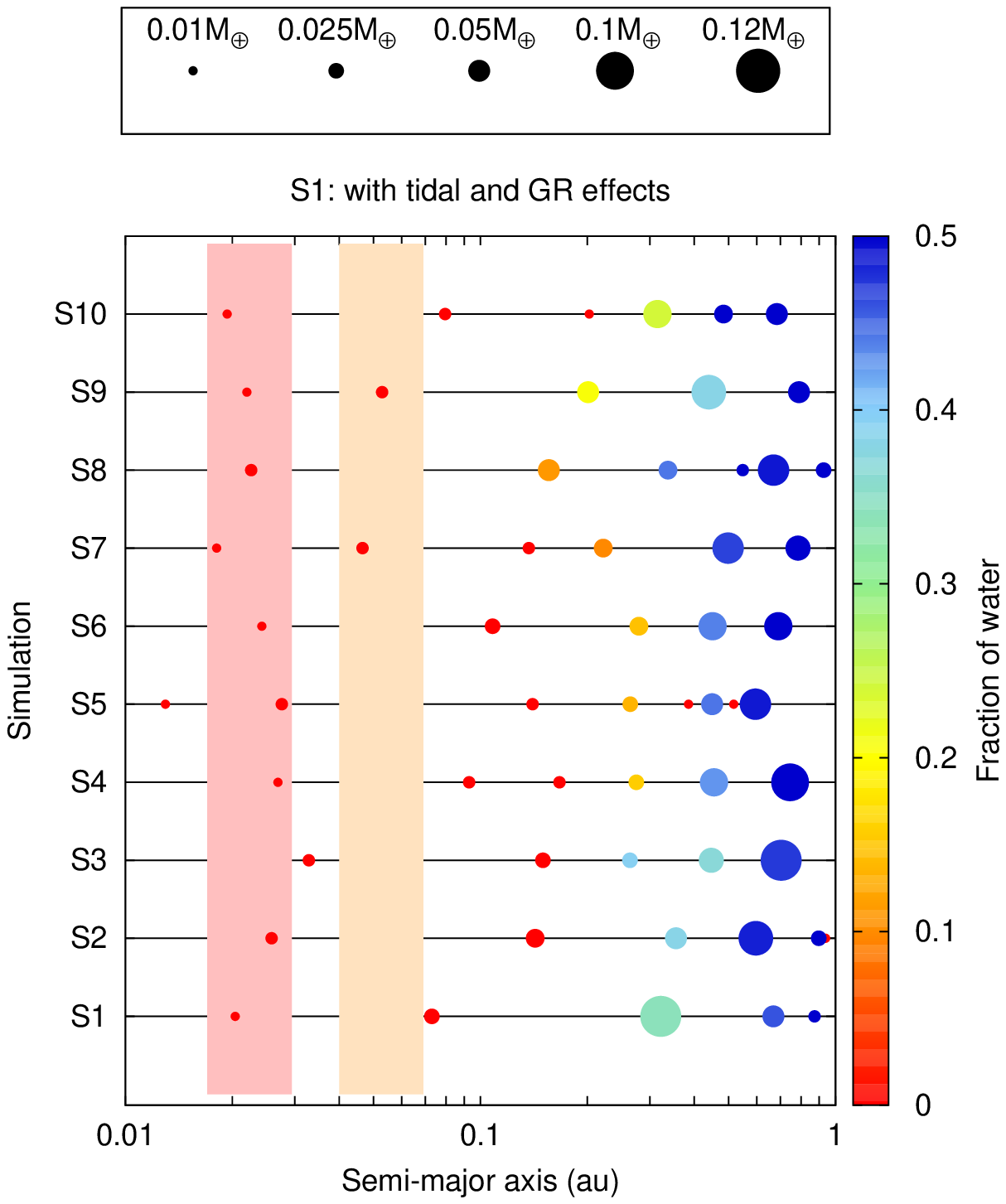}
    \includegraphics[width=0.4\textwidth]{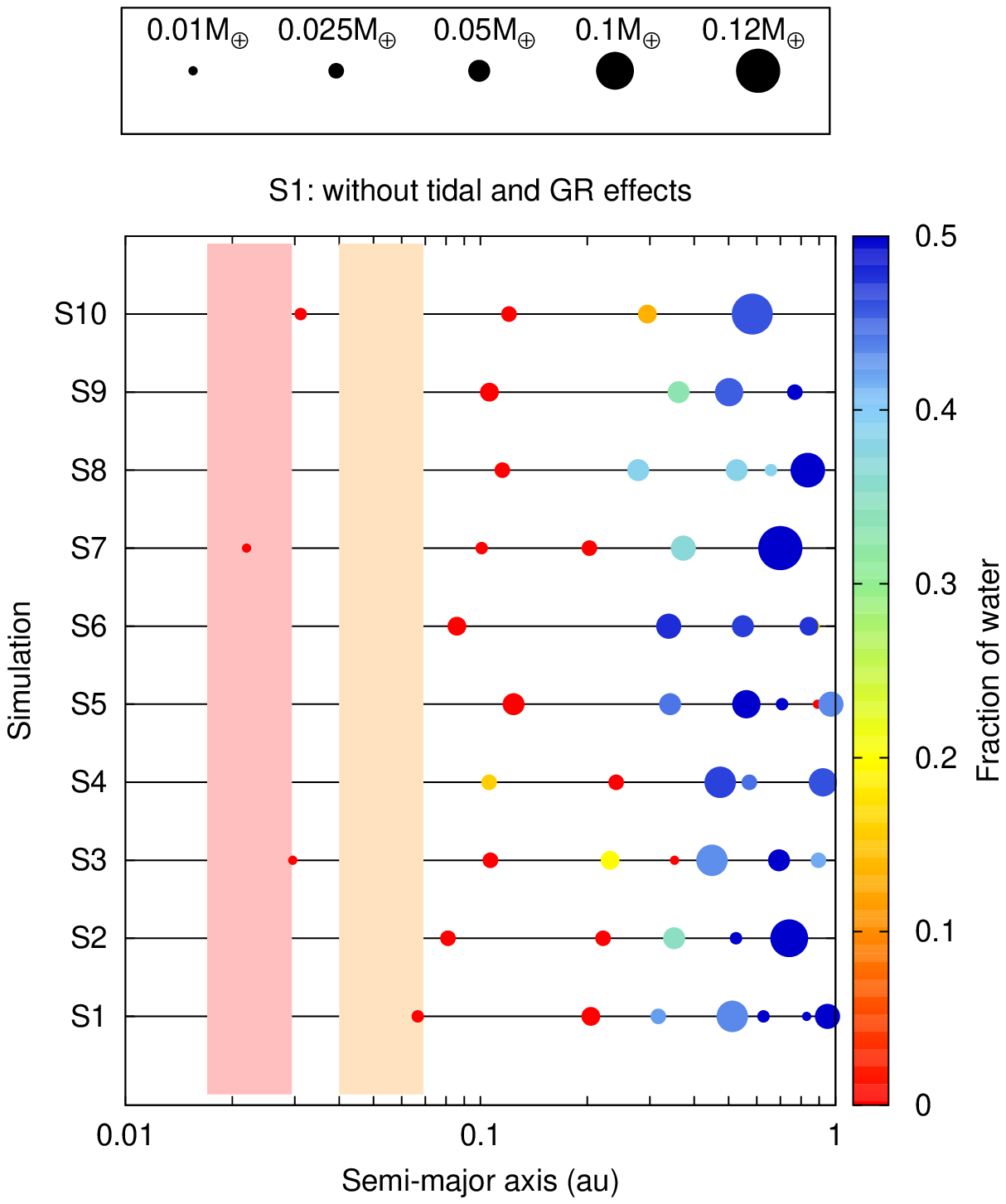}
        \caption{Distribution of the resulting planet locations in the region of study in 
        each simulation for scenario S1. Planets were distinguished by mass and fraction
        of water in mass. The masses of the resulting planets are shown at the top of each 
        graphic. The range is between 0.01~M$_\oplus$ and 0.12~M$_\oplus$ (this is approximately the mass of the $\text{Moon}$ and Mars, respectively). The 
        fraction of water in mass is presented in color-scale and is assigned to each body as 
        a percentage  between $0.01\%$ and $50\%$. The left panel represents the resulting 
        planets from simulations in which tidal and GR effects are included during the integration 
        time, while the right panel represents the planets from simulations that neglected these effects. The 
        cream band represent the IHZ at 100~Myr, and the pink band shows the IHZ at 1~Gyr.}
    \label{fig:Scenario1_architectures}
\end{figure*}

\begin{figure*}
\centering
    \includegraphics[width=0.4\textwidth]{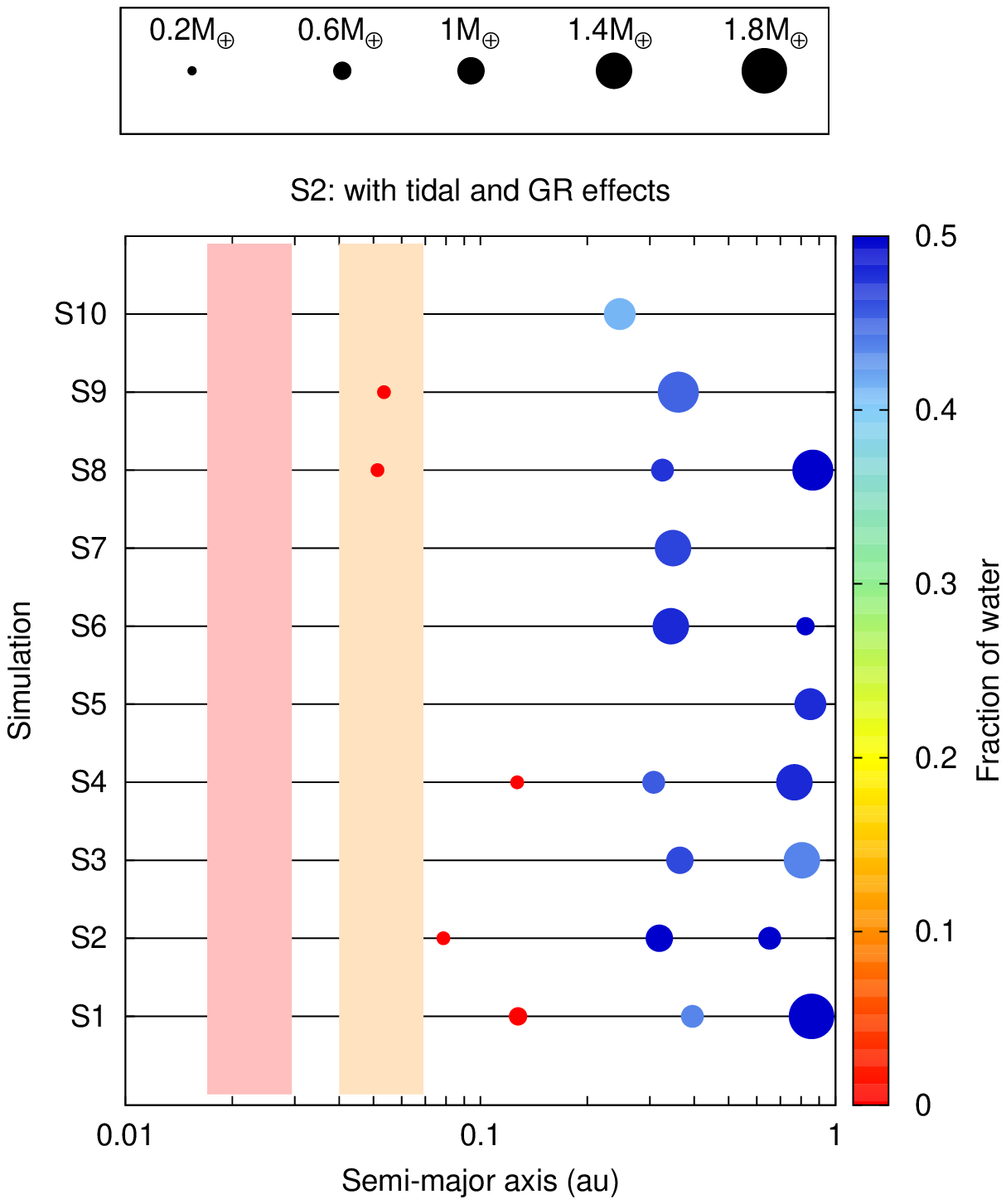}
    \includegraphics[width=0.4\textwidth]{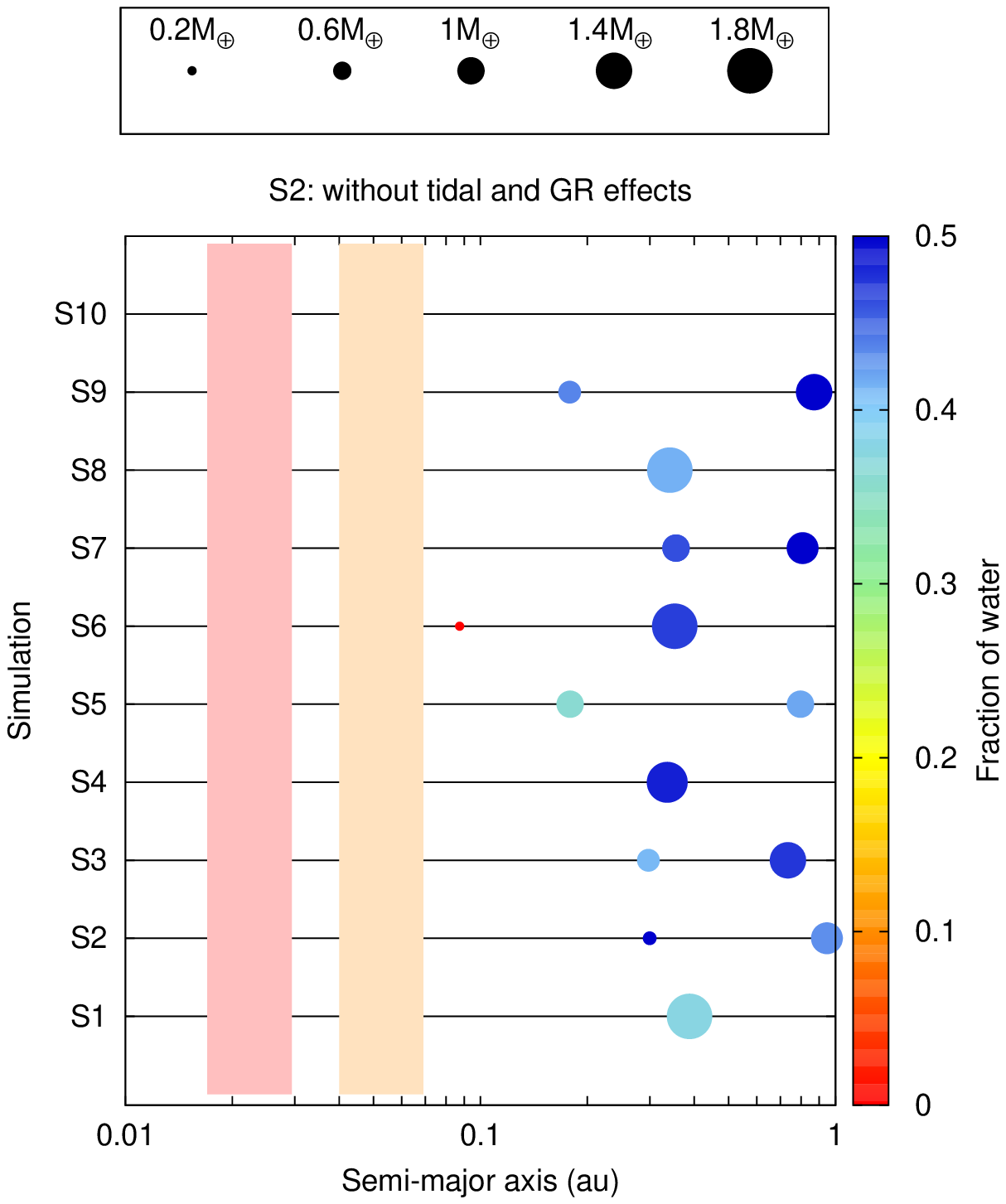}
    \caption{Distribution of the resulting planet locations in the region of study in each 
    simulation for scenario S2. The mass range of the resulting planets is between 0.2~M$_\oplus$ and 1.8~M$_\oplus$. The color characterization is the same as in Fig~\ref{fig:Scenario1_architectures}.}
  \label{fig:Scenario2_architectures}
\end{figure*}

Our simulations predict a diversity of final planetary system architectures 
at 100~Myr regarding all the simulations made in both scenarios. Fig.~\ref{fig:Scenario1_architectures} shows the final location of the 
resulting planetary systems of each simulation in scenario S1 for the 
final masses and fraction of water in mass. The planetary masses are between 0.01~M$_\oplus$
and 0.12~M$_\oplus$ (this is approximately the mass of the \emph{\textup{Moon}} 
and \emph{\textup{Mars,}} respectively) and the fraction of water in mass 
is between $0.01\%$ and $50\%$. The left panel shows the planetary architectures 
from simulations that included tidal and GR effects, and the right
panel presents their counterparts in the simulations that neglect these effects. 
In both panels, the IHZ of the system at 100~Myr   and at 1~Gyr overlap. 
In Fig.~\ref{fig:Scenario2_architectures} we show the resulting
architectures of the simulations for scenario S2.
The resulting planets have a range of masses between 
0.2~M$_\oplus$ and 1.8~M$_\oplus$ and a percentage of water in mass 
between $0.01\%$ and $50\%$. 

The main difference we found is the close-in planet
population that survived in the simulations that included
tidal and GR effects, which did not survive in the simulations
that neglected these effects. This becomes more relevant
in S1, where the protoplanetary embryos involved were an
order of magnitude less massive than in S2. 

\begin{figure}
\centering
  \includegraphics[width=8cm]{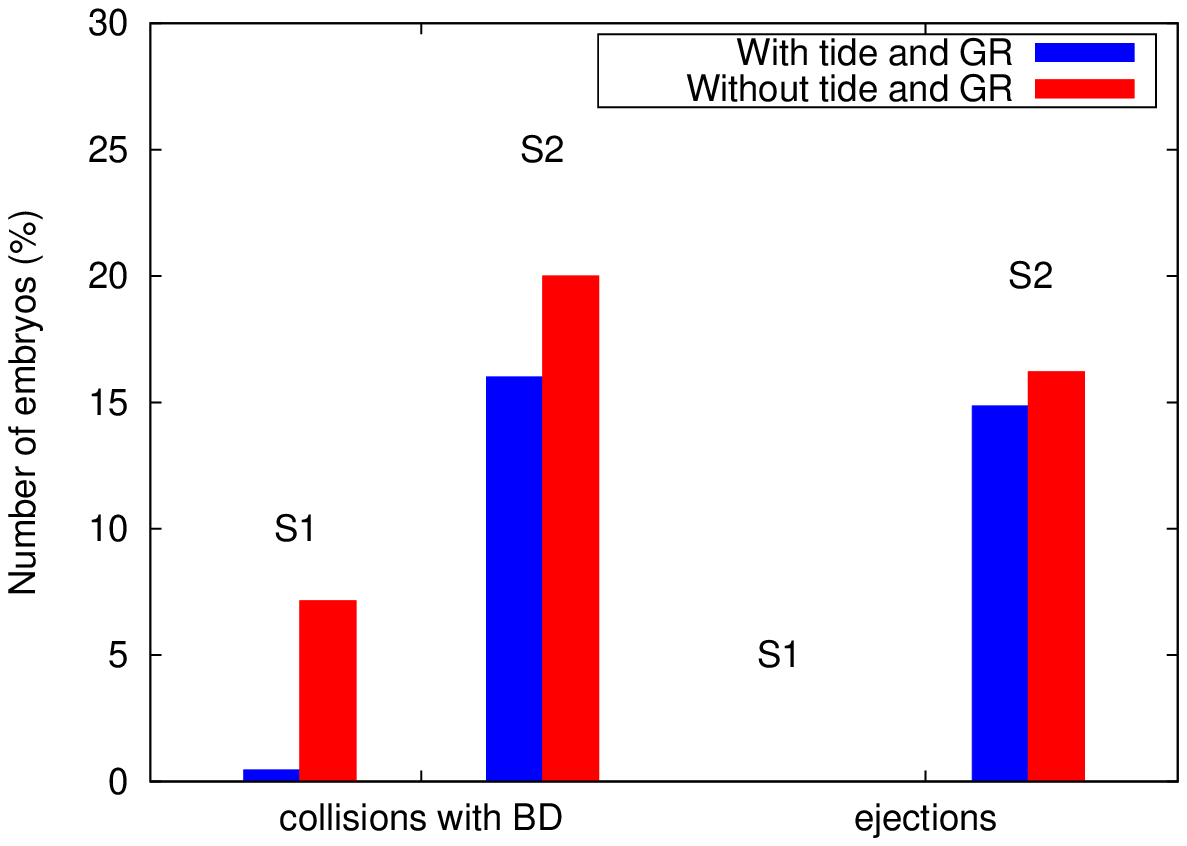}
  \caption{Percentage of embryos that collided or were ejected from the 
  system during the integration time of each scenario. Blue bars correspond 
  to embryos from simulations that included tidal and GR effects, and red bars represent embryos from simulations that neglected these effects.}
  \label{fig:colisiones}
\end{figure}

In simulations S2, the embryos involved suffered
stronger gravitational interactions between them than those 
from S1 because they are more massive bodies. Therefore they 
generate more excitation in the system, allowing some embryos 
to collide with the central object and to be ejected from the
system. These interactions became more relevant than tidal 
and GR effects for the population of very close-in bodies in S2. 
This is supported by the percentage of embryos that
collided with the central object or were ejected from the system
(reached $a > 100$~au), as shown in Fig.~\ref{fig:colisiones}.
The percentage is given over the initial amount of embryos 
in each scenario of work: 224 embryos in S1, and 74 embryos 
in S2. The number of
bodies that either collided with the central object 
or were ejected from the system in S2 is much higher 
than in S1 because the system is more highly excited. 
Moreover, the simulations that included tidal and GR 
effects reduced the collisions of embryos with the central
object and had almost no effect on the ejection of embryos
because at long distances from the central object, tidal
effects became irrelevant. 

\begin{figure}
    \centering
    \includegraphics[width=9cm,height=6cm]{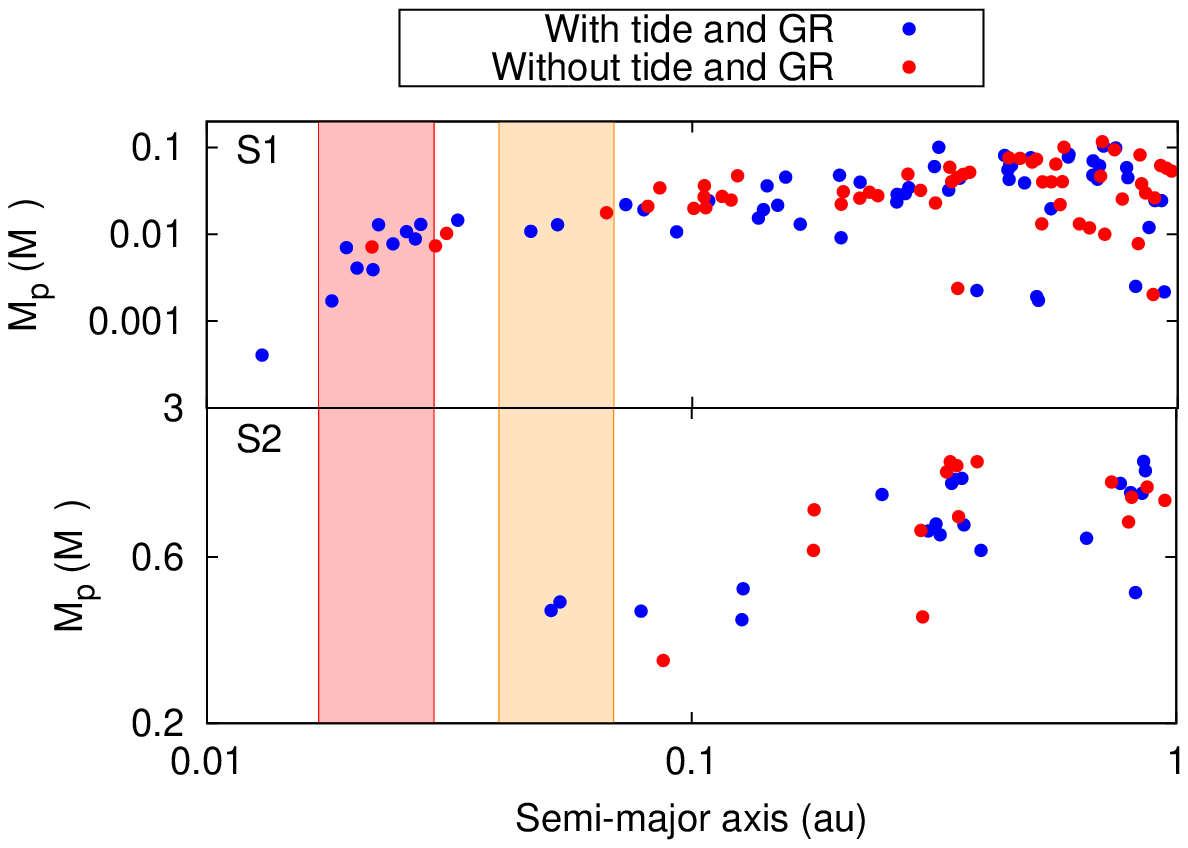}
    \caption{Distribution in mass of the resulting planets for their semimajor axis at 100~Myr in S1 (top panel) and S2 (bottom panel). Blue dots represent the planets from 
    simulations that included tidal and GR effects, while red dots represent those from
    simulations that neglected these effects. The pink band represents the location of the 
    IHZ at 1 Gyr, while the cream band represents its location at 100 Myr.}
    \label{fig:masas_vs_a}
\end{figure}

By compacting all the resulting planetary systems in
Fig.~\ref{fig:masas_vs_a}, we represent their distribution 
in $a$ for their masses for S1  and S2. The IHZ at 100~Myr and at 1~Gyr overlap as well. The 
surviving population of close-in bodies has low masses and
mainly results from simulations with tidal and GR effects
from S1. This shows that the relevance of tidal and GR effects also depends on the mass of the bodies that are involved in the simulations.

\begin{figure}
    \centering
    \includegraphics[width=9cm,height=6cm]{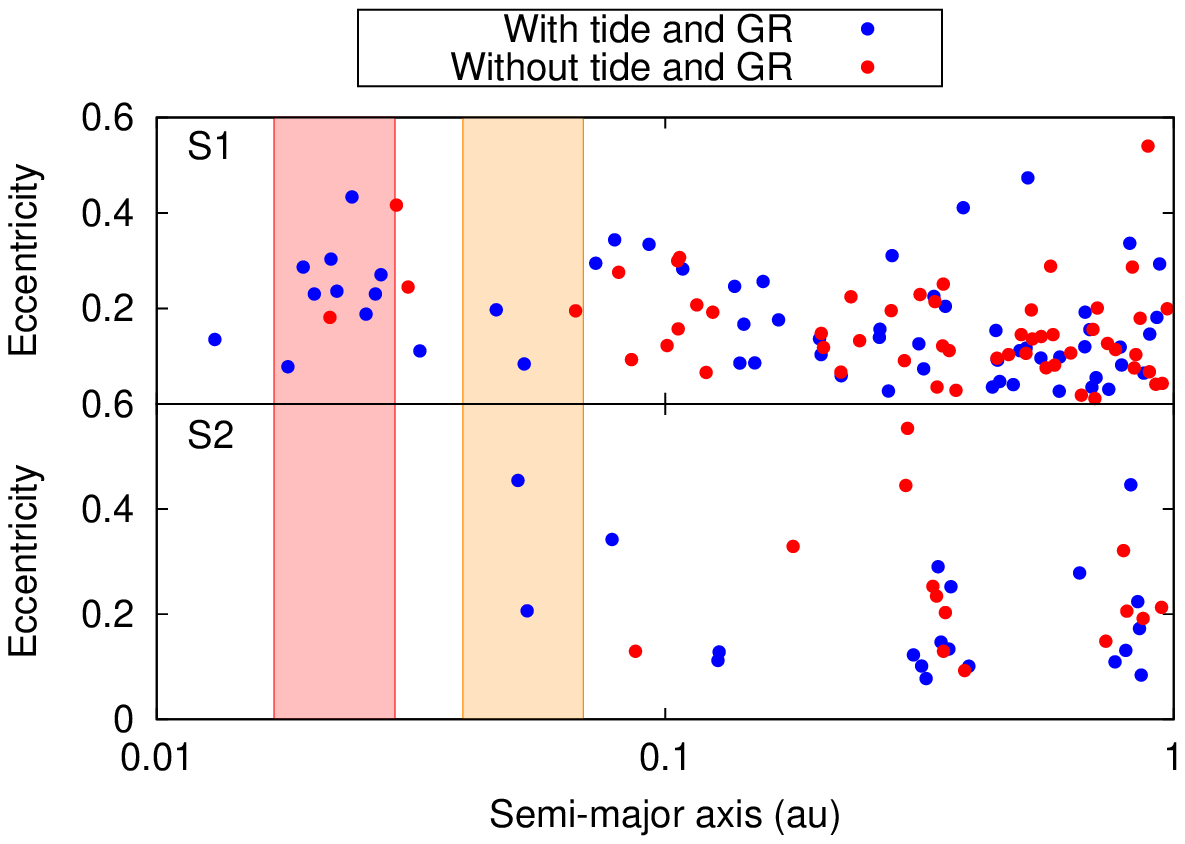}
    \caption{Orbital distribution of the resulting planets regarding their location in the system and eccentricity in S1 (top panel) and S2 (bottom panel). Colors are as in Figure \ref{fig:masas_vs_a}.}
    \label{fig:avse}
\end{figure}

The Fig. \ref{fig:avse} shows the eccentricities of the resulting 
planets as a function of their semimajor axis
for S1 and S2. Low-mass and close-in planets that survived while 
external effects were included appear to have
eccentricities values greater than zero. This is because the many gravitational
interactions between embryos produce excitation in their 
orbital parameters, and the timescale for eccentricity 
damping is far longer ($\text{approximately some billion years}$) than the 
integration time of our simulations. As we discussed in
previous sections, the tidal effects added in our
simulations affect the distribution of eccentricities and
the semimajor axis of the resulting planets, but the long decay
timescales prevent us from seeing the damping in $e$ at this point. Nevertheless, 
the $e$ damping will be efficient by the 
time the central object reaches 1~Gyr for the planets
that remain located close in to the central object, where 
the IHZ will be located by that time.

Our results strongly suggest that a formation scenario that includes 
tidal and GR effects is more
realistic for planet formation at the substellar 
mass limit. Although GR 
corrections are relevant during planet formation, the tidal effects are mor important to map 
more realistic orbits and therefore more realistic 
encounters between embryos. These effects play a 
primary role in the survival of an in situ population. However, 
when the masses of the bodies involved 
increase (like in S2), tidal effects became less relevant 
than the gravitational interactions between them 
(see Section~\ref{sec:discussions}).

\subsection{Water mass fraction}
\label{sec:water_planets}

The two scenarios show a diversity in the fraction of water in mass, but the resulting planets inside
$a<0.1$~au always conserved this initial fraction of water in mass. We assumed this to be $0.01\%$ of the mass of the embryos that are located
inside the snow line at the beginning of the simulations. 

For S1, planets at $a>0.1$~au present a range in
percentage of water in mass that is between $10\%$ and
$35\%$ for planets with a semimajor axis $0.1<a/\textrm{au}<0.42$, 
and it is between $35\%$ and $50\%$ for planets with
$0.42<a/\textrm{au}<1$. The outer water-rich planets maintain
their high content of water in mass, and an intermediate
population of water-rich resulting planets appears close 
to the location of the snow line.

For S2, planets at $a>0.1$~au present a high mass 
percentage of water, between $35\%$ and $50\%$.
Embryos located outside the snow line either have suffered impacts of 
other water-rich bodies or have been ejected from the system. No intermediate water-rich population as in S1 evolved. In order to explain the origin of the resulting distribution
of water of the surviving planets in our region of study, 
in the next section we analyze their whole collisional 
history.

\begin{figure*}
    \centering
    \includegraphics[width=0.65\textwidth]{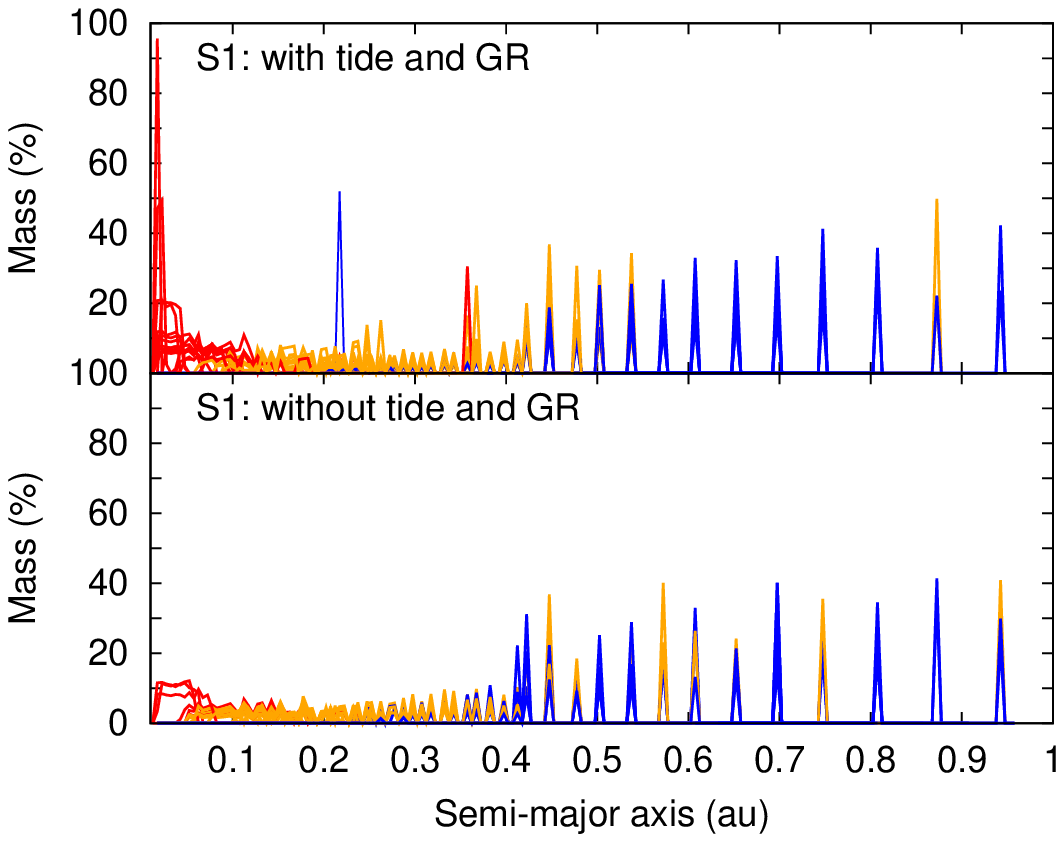}
    \caption{Collisional history of the resulting planets from the simulations in S1 in which tidal and GR effects are included (top panel) and in which these effects are neglected (bottom panel). Each jump in semimajor axis indicates the initial location of the embryo that collided with the resulting planets that increased their masses by a given percentage after the perfect merger. The red lines indicate the history of the resulting planets that are finally located at $a < 0.1$~au, the orange lines show planets located at $0.1 < a/\mathrm{au} < 0.42,$ and the blue lines represent planets located at $0.42 < a/\mathrm{au} < 1$.}
    \label{fig:historia_colisional}
\end{figure*}

\begin{figure*}
    \centering
    \includegraphics[width=0.65\textwidth]{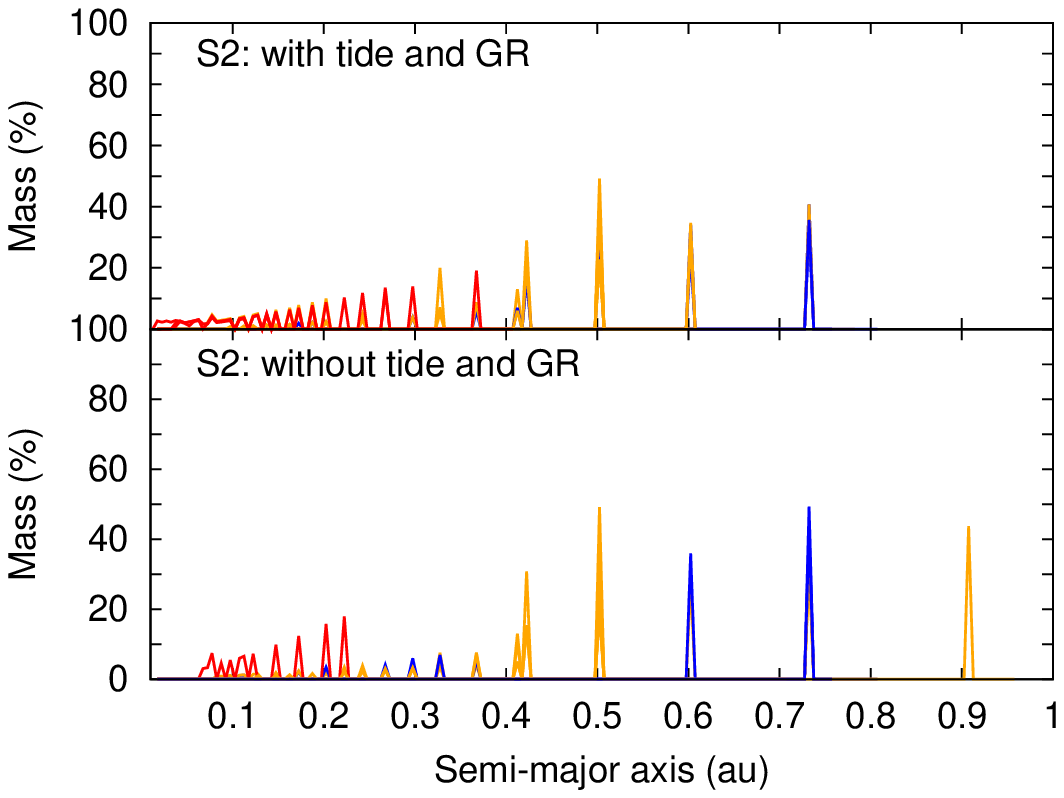}
    \caption{Collisional history of the resulting planets from the S2 simulations in which tidal and general relativistic effects are included (top panel) and in which these effects are neglected (bottom panel). Colors are the same as in Fig.~\ref{fig:historia_colisional}.}
    \label{fig:historia_colisional2}
\end{figure*}

\subsection{Collisional history}
\label{sec:sollisions}

During the first 100~Myr of planet formation that we studied
in our simulations, embryos had gravitational interactions in the form of encounters and collisions among them. From 
the initial location in the system, each embryo can interact 
with others that have different orbital and physical 
parameters in its gravitational influence zone. The 
$N$-body code treats all collisions as perfectly inelastic. After two bodies collide, the resulting body therefore is 
a merger of the initial two.  

From the previous analysis of the final distributions 
of the resulting planets and their final fraction of water in
mass, we can distinguish different subregions in which 
the collisional history of the resulting planets can be studied: an inner
region of planets that are finally located at $a < 0.1$~au, an
intermediate region, between $0.1<a/\mathrm{au}<0.42,$ and an 
outer region beyond the snow line, between $0.42<a/\mathrm{au}<1$. 

Fig.~\ref{fig:historia_colisional} shows the collisional history of all the resulting planets of S1 when tidal and GR effects are included and when these effects are neglected. Each peak in the lines represents the initial location of the embryo that collided with the resulting planet and the percentage of mass that it added to the planet after the perfect collision. In the inner region the resulting planets collided with embryos that were initially located at $a < 0.35$~au, all located inside the snow line, which means that they preserve the initial fraction of water in mass. In the intermediate region, planets accreted embryos that were initially located between $0.05 < a/\mathrm{au} < 0.88$  and between $0.05 < a/\mathrm{au} < 0.95,$  which explains the intermediate range of water in mass after collision with embryos inside and outside the snow line. Finally, in the outer region, planets collided with embryos that were initially located between $0.2 < a/\mathrm{au} < 0.95$ and $0.22 < a/\mathrm{au} < 0.95$ . In this outer region, even though the resulting planets suffered collisions with embryos that were distributed in a wide range of the system, only a few planets that were located inside the snow line collided with these planets and thus maintained a huge fraction of water in mass. It is important to remark that the close-in population located at $a < 0.1$~au did not collide with other embryos beyond this semimajor axis when tidal effects were included in the simulations.

On the other hand, Fig.~\ref{fig:historia_colisional2} shows the collisional history of the resulting planets of S2 when tidal and GR effects are included and when these effects are neglected. In this case, in the inner-region planets collided with embryos that were initially located at $a < 0.36$~au and $a < 0.22$~au. In the intermediate region, planets accreted embryos that were initially located between $0.06 < a/\mathrm{au} < 0.72$  and $0.07 < a/\mathrm{au} < 0.91$ . Finally, in the outer region, planets collided with embryos that were initially located at $0.17 < a/\mathrm{au} < 0.72$  and $0.097 < a/\mathrm{au} < 0.72$. In S2, planets in the intermediate and outer region accreted material from a similar region, with a few embryos initially located inside the snow line. Thus all these planets retained a huge amount of water in mass. In this case, the three planets that survived with at $a < 0.1$~au did not collide with other embryos beyond this semimajor axis as happened in S1 when tidal effects were included in the simulations.

The collisional history of the resulting planets explains their final masses and the fraction of water in mass. Moreover, it allows us to conclude that the close-in surviving population that was most affected by tidal effects only suffered collisions with the embryos that were initially located close in.

\subsection{Close-in population: potentially habitable planets}
\label{sec:HZ_planets}

We focus our analysis on the close-in bodies that survived in the simulated planetary systems. We gave physical and orbital parameters of those bodies candidates to be potentially habitable planets.

\subsubsection{Characterization}

Inside the location of the IHZ at 100~Myr (final time of simulations), two planets in S1 remained  when tidal and GR effects were included in the 
simulations and only one planet remained when these effects were excluded from the simulations 
regarding their semimajor axis. On the other hand, in S2 two planets remained in the 
IHZ when external effects were included and no candidate survived when these effects were excluded from the simulations. Moreover, when we consider the value of the eccentricity and 
calculate the periastron distance (q) and apastro distance (Q) of these planets, only one planet remained inside the 
IHZ in a, q, and Q in S1 and 1 in S2 when tidal and general relativistic 
effects were considered in the simulations.

In Section \ref{sec:habzone} we discussed the behavior
of the IHZ around very low mass stars that are located close
to the star and evolve toward a smaller radius as the star
evolves with time.  We therefore extended our analysis to bodies that ended up closer in to the central object, 
in particular, inside the location of the IHZ at 1~Gyr. The
S1 alone generated such a close-in body 
population: nine planets when
tidal and GR effects were incldued, and only one planet when these effects were neglected in the simulations regarding their semimajor axis. 
Even though we do not have the orbital parameters at 1~Gyr, it is expected that the eccentricities of these planets are small enough for them to remain in the IHZ at 1~Gyr in S1 (see Section~\ref{sec:numodel}). In Table \ref{tab:ZH_info} we present
some physical parameters of the planets that remained in the IHZ at 
100~Myr or at 1~Gyr in both scenarios. When tidal and GR effects are included in 
the simulations, S1 is the most favorable scenario to allow these candidates of 
potentially habitable planets (see Section~\ref{sec:discussions} for further discussion.)

\begin{table}
    \centering
    \begin{tabular}{c|c|c|c|c|c|c}
        \hline
         Scenario & Embryo &    a   & $e$ & M              & $H_2O$ & IHZ\\
             &        & [$au$] &     & [M$_\oplus$]        &  [\%]  &    \\
         \hline 
         S1t  &  39 & 0.020 & 0.23  & 0.004 & 0.01 & b \\
         S1t  & 132 & 0.025 & 0.18  & 0.010 & 0.01 & b \\
         S1t  & 113 & 0.026 & 0.23  & 0.008 & 0.01 & b \\
         S1t  & 113 & 0.027 & 0.27  & 0.013 & 0.01 & b \\
         S1t  & 132 & 0.024 & 0.43  & 0.007 & 0.01 & b \\
         S1t  &  33 & 0.018 & 0.07  & 0.001 & 0.01 & b \\
         S1t  & 126 & 0.046 & 0.19  & 0.010 & 0.01 & a \\
         S1t  & 115 & 0.022 & 0.23  & 0.013 & 0.01 & b \\
         S1t  &  63 & 0.022 & 0.30  & 0.006 & 0.01 & b \\
         S1t  & 142 & 0.052 & 0.08  & 0.013 & 0.01 & a \\
         S1t  &  62 & 0.019 & 0.28  & 0.007 & 0.01 & b \\
         S1wt & 163 & 0.066 & 0.195 & 0.017 & 0.01 & a \\
         S1wt &  90 & 0.022 & 0.18  & 0.007 & 0.01 & b \\
         S2t  &  59 & 0.051 & 0.45  & 0.337 & 0.01 & a \\
         S2t  &  39 & 0.053 & 0.20  & 0.370 & 0.01 & a \\
        \hline
    \end{tabular}
    \caption{Potentially habitable planets from both scenarios when tidal and GR effects are included in the simulations (S1t and S2t) 
    and when these effects are neglected in the case of S1 (S1wt). The initial 
    numbers of embryos that become the resulting planet are listed in Col. 2 with 
    their respective final  semimajor axes, eccentricity, mass, and 
    percentage of water in mass. In their final locations, the planets are 
    located in the IHZ at 100~Myr (IHZ = a) or in the IHZ at 1~Gyr (IHZ = b).}
    \label{tab:ZH_info}
\end{table}

\subsubsection{Mass accretion history}
\label{sec:colid_history_IHZ}

\begin{figure}
    \centering
    \includegraphics[width=8cm]{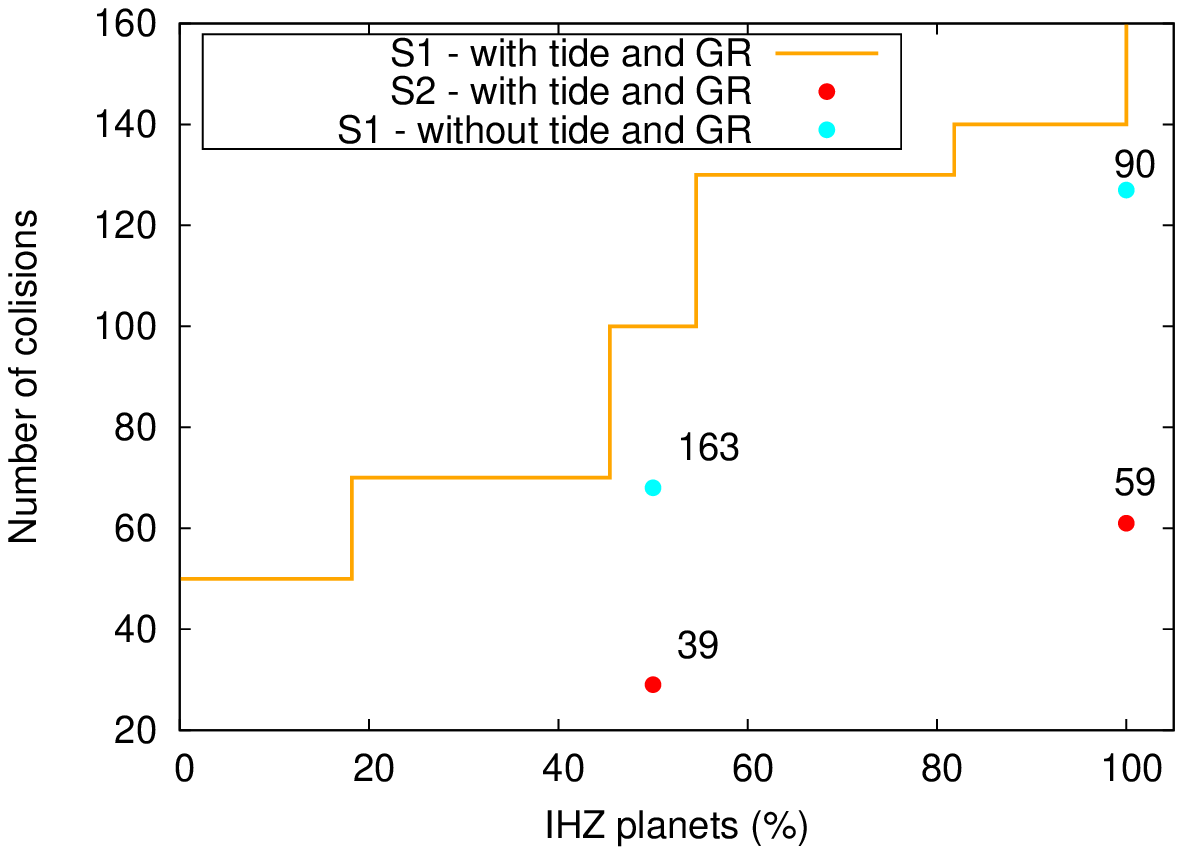}
    \caption{Cumulative collisions between the resulting planets 
    that survived inside the IHZ at 100~Myr and at 1~Gyr for S1 
    with tide and GR (orange line), S1 without considering tide or 
    GR (cyan dots), and S2 with tide and GR.}
    \label{fig:Scenario1_colissions}
\end{figure}

All the close-in population that we consider as candidates for potentially habitable 
planets had many collisions during the integration time of our 
simulations. All of them were targets of many impacts. We show in
Fig.~\ref{fig:Scenario1_colissions} the number of collisions of all the IHZ 
candidates at 100~Myr and 1~Gyr. In S1, all the resulting planets received more than 50 impacts and a maximum of 160 impacts, and 
$50\%$ of the total received more than 100 impacts when tidal and GR effects were included, while
the planets received 70 and 127  impacts when these effects were neglected. In S2, one of the planets received 
29 impacts and the other almost 70 impacts when tidal and GR effects were included. Each impact corresponds to a collision 
with another embryo of the simulation. In S1 more impacts are allowed because the total
number of embryos is much higher ($224$ in total) than in S2 ($74$ in total). In any case, 
all the candidate planets collided with between 25$\%$ and $50\%$ of the total number of embryos during 
the first 100~Myr of their evolution. 

\begin{figure}
    \centering
    \includegraphics[width=8cm]{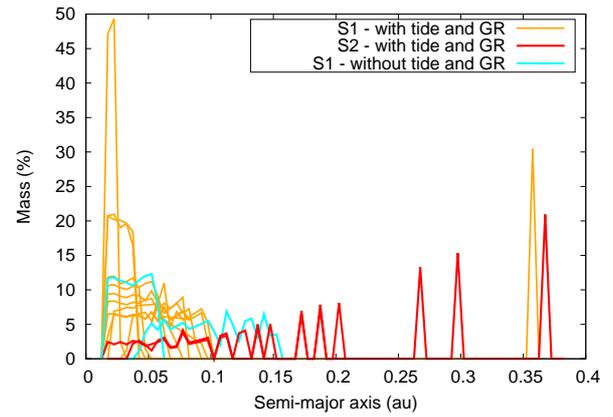}
    \caption{Initial location of each embryo that collided with planets that survived 
    inside the IHZ at 100~Myr or 1~Gyr related to the percentage of mass 
    that the candidate planet obtained after each collision in S1 with tide and 
    GR (orange curves), in S2 without tide and GR (cyan curves), and in S2 with 
    tide and GR (red curves).}
    \label{fig:planets_impactors}
\end{figure}

Table~\ref{tab:ZH_info} shows that all the planets inside the 
IHZ conserved their initial fraction of water in mass because all the impacts that they suffered came 
from embryos that were located inside the snow line of the system. Fig.~\ref{fig:planets_impactors} shows the location on the semimajor axis of 
each embryo that collided with one IHZ candidate, related to the percentage 
of mass that the candidate obtained after each collision in S1 with tide and 
GR, in S2 without tide and GR, and in S2 with 
tide and GR. This figure is a zoom of Fig.~\ref{fig:historia_colisional} and Fig.~\ref{fig:historia_colisional2} for the planets located in the two determined IHZ.

\begin{figure}
    \centering
    \includegraphics[width=0.45\textwidth]{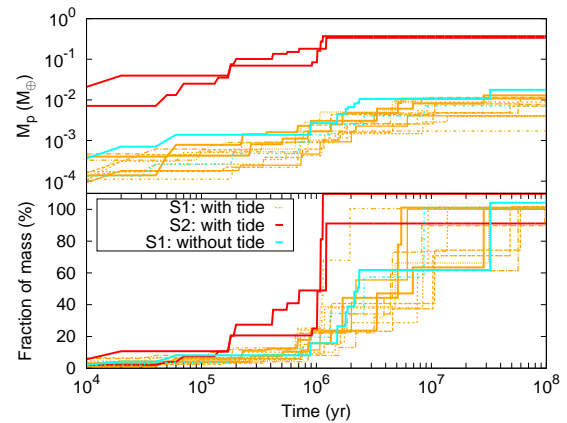}
    \caption{Evolution of the mass (top panel) and its fraction with respect 
    to the final mass (bottom panel) of the resulting planets that survived 
    inside the IHZ at 100~Myr (solid lines) and 1~Gyr (dotted lines) in S1 
    with tide and GR (orange curves), in S2 without tide and GR (cyan curves), 
    and in S2 with tide and GR (red curves).}
    \label{fig:planets_masses}
\end{figure}

The impacts that each body suffered during the integration time always produced 
an increase in mass because the N-body code we used to develop the simulations 
considers all collisions as completely inelastic, so that every time a collision 
between embryos occurs, it ends in a perfect merger (see Section~\ref{sec:discussions}). 
In Fig~\ref{fig:planets_masses} we show the evolution in mass of each IHZ candidate 
planet and its fraction of mass. Each step in the curves represents the 
mass or fraction of mass, respectively, that is gained by the IHZ candidate planet after each 
collision. There is no difference between candidate planets that were finally 
located in the IHZ at 100~Myr and 1~Gyr in each scenario with respect to 
the mass accretion history.\\

\section{Conclusions and discussions}
\label{sec:discussions}

We studied the rocky planet formation around a star close to the substellar 
mass limit using $N$-body simulations that included tidal and GR 
effects and did not include the effect of gas in the disk. Our aim was to evaluate the 
relevance of tidal effects following the equilibrium tidal model 
during the formation and evolution of the system and to improve the accuracy in the 
calculation of the orbit of the protoplanetary embryos by considering GR effects.

The equilibrium tide model we used is based on the assumption 
that when a star suffers tidal disturbance from a companion body, it instantly adjusts 
to hydrostatic equilibrium \citep{Darwin1879}. 
A more general approach must include
the dynamical tide model for a more reliable
description of the very high eccentricity orbits, 
as shown by \citet{Ivanov2011} and
references therein. As shown in Figure 
\ref{fig:colisiones}, in simulations that do
not include tides, 7\% and 20\% of the embryos
collide with the central star in S1 
and S2, respectively, because of their high eccentricity. When tides are included through the equilibrium
model, these fractions change to 
1\% and 16\%, respectively.
New simulations that include the dynamical tide 
model are needed in order to explore the possible
change in these fractions, which we expect to
occur toward lower values because the 
tidal damping produced by this model is stronger.





Using a different model for close-in bodies, \citet{Makarov2013} found 
non-pseudo-synchronization in the rotational periods for terrestrial planets and moons. 
In our case, we adopted the pseudo-synchronization to maintain consistency with 
the \cite{Hut1981} model that we adopted for the $N$-body simulations because the hypothesis of
pseudo-synchronization is a direct consequence of the constant time-lag model
\citep{Darwin1879, Hut1981, Eggleton1998}. A comparative analysis of our results with 
those from other treatments as well as the self-consistent inclusion of the rotational 
evolution of embryos is beyond the goals of this work.

Because of the current uncertainties in the determination 
of disk masses, the simulations were performed for two different scenarios S1 and S2,
which basically differ in the initial mass of solid material in the disk, 
$\sim 3$~M$_\oplus$ and $\sim 30$~M$_\oplus$ , respectively. These values roughly represent 
the corresponding upper and lower limits of the disk mass for stars
close to the substellar mass limit.

The resulting planets have masses between $0.01 < m/\mathrm{M_\oplus} < 0.12$
in scenario S1 and $0.2 < m/\mathrm{M_\oplus} < 1.8$ in scenario S2. Even though we used 
lower values for the disk mass than those from \citet{Payne2007}, we found the same 
correlation between the resulting planet masses and the initial amount of solid material 
in the disk. When the disk mass increases, more massive planets 
could be formed.

When tidal and general relativistic effects are included, a close-in planetary 
population located at $a < 0.07$~au in scenario S1 survived in all the simulations, while in the more massive 
scenario S2, embryos suffered stronger gravitational interaction and the formation of this 
close-in population occurred only in two of the ten simulations.  S1 therefore is the most 
favorable scenario for generating close-in planets. Then, we found that tidal and general 
relativistic effects are relevant during the formation and evolution of rocky planets 
around an object at the substellar mass limit, in particular when the protoplanetary 
embryos involved are low-mass bodies. Our work together with the model developed by 
\citet{Bolmont2011,Bolmont2013,Bolmont2015} shows that tidal effects in both the 
formation and later evolution of such systems are required.

The close-in population resulted from a large number of collisions among the protoplanetary 
embryos, which was treated in our simulations as perfectly inelastic collisions. This gives upper 
limits on the final mass value and water content for the resulting planets. This shows why 
it is necessary to reproduce the simulations using an $N$-body code that includes 
fragmentation during collisions, which can decrease the final masses of the 
resulting planets considerably, as shown by \citet{Chambers2013,Dugaro2019}.  

The close-in population we found is of particular interest because it is located inside the 
evolving IHZ of the system. We classified a set of 15 close-in bodies as candidates to 
potentially habitable planets based on their semimajor axes and eccentricities.
A complete analysis of their probability of being habitable planets considering
additional constraints \citep[e.g.,][]{Martin2006} is beyond the scope of this work. 

Our model presents an important improvement in the scenario of rocky planet formation 
at the substellar mass limit by including tidal and general relativistic effects. 
We stress that even tough \citet{Coleman2019} did not incorporate these effects in 
their simulations of planet formation around Trappist 1, they showed the relevance of 
considering the gas component of the disk during the first 1~Myr. A more 
realistic simulation of this scenario of planet formation must therefore clearly include all these effects. 
A new set of such $N$-body simulations considering low-mass stars and BDs as central objects with different 
masses is currently ongoing.

We conclude that tidal and GR effects are relevant during rocky planet formation 
at the substellar mass limit because they allow the survival of close-in bodies that are located 
inside the IHZ. This supports the hypothesis that these systems are important candidates for future searches of 
life in the solar neighborhood.

\begin{acknowledgements}
This work was partially financed by Agencia Nacional de
Promoci\'on Cient\'ifica y Tecnol\'ogica (ANPCyT) through
PICT 201-0505, and by Universidad Nacional de La Plata
(UNLP) through PID G144. We acknowledge the financial
support by Facultad de Ciencias Astron\'omicas y
Geof\'isicas de La Plata (FCAGLP) and Instituto de
Astrof\'isica de La Plata (IALP) for extensive use 
of their computing facilities. We specially appreciate
the kind support and advice from 
Bolivia Cuevas-Otahola (INAOE, M\'exico) during the
numerical simulations and Adri\'an Rodr\'iguez Colucci 
(UFRJ, Brazil) for his valuable comments. 
We thank the editor Beno\^it Noyelles and the 
anonymous referee for very
valuable suggestions which helped improve the 
presentation of our results.
\end{acknowledgements}

%
\bibliographystyle{aa} 
\bibliography{Referencias} 
%

\begin{appendix}
\section{Snow-line location in the substellar regime}
\label{sec:Apendix}

The snow-line location corresponds to the radial distance to the central object where water 
condenses. This occurs when the partial pressure of the protoplanetary disk exceeds the 
saturation pressure. The exact temperature where this occurs depends on the physical structure
of the disk and on the relative abundances of elements, but is expected to be in a range between
$140<T/\textrm{K}<170$. To determine the snow-line location in a system with a substellar 
object of a mass 0.08~M$_\odot$ as a primary object, we took the same temperature profile 
of a disk that radiates as a blackbody as \citet{Chiang1997},






\begin{equation}
  T(r)=T_\star\left(\frac{\alpha}{2}\right)^{\frac{1}{4}}\left(\frac{r}{R_{\star}}\right)^{-\frac{1}{2}}.
\label{eq:temp}
\end{equation}

 The parameter $\alpha$ represents the glazing angle at which the starlight strikes the disk. \citet{Chiang1997}  considered vertical hidrostatic equilibrium and derived the expression 

\begin{equation}
\alpha = \frac{0.4R_\star}{a} + a \frac{d}{da}\left(\frac{H}{a}\right),    
\label{eq:alpha}
\end{equation}
where the parameter $H$ represents the height of the visible photosphere above the disk mid-plane, and the factor $(H/a)$ can be expressed by

\begin{equation}
\frac{H}{a} = 4T_\star^{\frac{4}{7}}R_{\star}^{\frac{2}{7}}\left(\frac{k}{GM_\star\mu_\textrm{g}}\right)^\frac{4}{7}a^{\frac{2}{7}},    
\label{eq:Ha}
\end{equation}
where $k$ is the Boltzman constant and $\mu_{\textrm{g}}$ is the mass of the molecular hydrogen. 

By replacing Eq. \eqref{eq:Ha} in Eq. \eqref{eq:alpha}, we estimated a value of $\alpha$. We considered then that the snow line is located where the disk reaches a temperature of $T(r = r_{\mathrm{ice}}) = T_{\mathrm{ice}} = 140 K$. Thus we could estimated the location of the snow line by 

\begin{equation}
r_{\textrm{ice}} = R_\star \left(\frac{T_\star}{T_{\textrm{ice}}}
\right)^2\left(\frac{\alpha}{2}\right)^{\frac{1}{2}}.
\label{eq:snow}
\end{equation}

We estimated the location of the snow line at $r_{\mathrm{ice}}=0.42$ au, when the BD is 1~Myr old, that is, the initial time of our simulations. We considered this value fixed in all the simulations. Because we know that $T_\star$ and $R_\star$ evolve with time, the location of the snow line would also evolve and continuously approach the star. When we consider that the location of the snow line evolves with time, it might have important consequences for the final amount of water in mass of the resulting planets of the simulations. However, this treatment is beyond the scope of this work. 




\end{appendix}

\end{document}